\documentclass[aps,pre,preprint,showpacs,superscriptaddress]{revtex4}
\usepackage{amsmath}
\usepackage{epsfig}

\preprint{APS/123-QED}

\newfont{\logo}{logo10}

\newcommand{\bea}{\begin{eqnarray}}
\newcommand{\eea}{\end{eqnarray}}
\newcommand{\ds}{\displaystyle}

\begin{document}
\bibliographystyle{revtex}
\title{Bright-dark solitons and their collisions in mixed N-coupled nonlinear Schr{\"o}dinger equations} 

\author{M. Vijayajayanthi\footnote{ e-mail: jayanthi@cnld.bdu.ac.in}}
\affiliation{Centre for Nonlinear Dynamics, School of Physics, Bharathidasan University, Tiruchirapalli--620 024, India\\} 
\author{T. Kanna\footnote{ e-mail: tkans@rediffmail.com}} 
\affiliation{Department of Physics,
 \\
Bishop Heber College,Tiruchirapalli--620 017, India}
\author{M. Lakshmanan \footnote{ Corresponding author e-mail: lakshman@cnld.bdu.ac.in}}
\affiliation{Centre for Nonlinear Dynamics, School of Physics, Bharathidasan University, Tiruchirapalli--620 024, India\\}      
      
\begin{abstract}
Mixed type (bright-dark) soliton solutions of the integrable N-coupled nonlinear Schr{\"o}dinger (CNLS) equations with mixed signs of focusing and defocusing type nonlinearity coefficients are obtained by using Hirota's bilinearization method.  Generally, for the mixed N-CNLS equations the bright and dark solitons can be split up in $(N-1)$ ways.  By analysing the collision dynamics of these coupled bright and dark solitons systematically we point out that for $N>2$, if the bright solitons appear in at least two components, non-trivial effects like onset of intensity redistribution, amplitude dependent phase-shift and change in relative separation distance take place in the bright solitons during collision.  However their counterparts, the dark solitons, undergo elastic collision but experience the same amplitude dependent phase-shift as that of bright solitons.  Thus in the mixed CNLS system there co-exist shape changing collision of bright solitons and elastic collision of dark solitons with amplitude dependent phase-shift, thereby influencing each other mutually in an intricate way.
\end{abstract}

\maketitle
\section{INTRODUCTION}
Solitons in coupled nonlinear Schr{\"o}dinger (CNLS) equations have been the subject of intense studies due to their intriguing collision properties and their robustness  against external perturbations.  The study of physical and mathematical aspects of CNLS equations is of considerable current interest as these equations arise in diverse areas of science like nonlinear optics, optical communication, biophysics, Bose-Einstein condensates (BECs) and plasma physics \cite{ref1,ref2,ref3,ref4,ref5,ref6}.  In the context of nonlinear optics solitons arising as a result of competing focusing type nonlinearity and anomalous dispersion (diffraction) of pulse (beam) are called bright solitons \cite{ref3} as they are well localized structures of light.  The dark solitons resulting due to compensation between defocusing type nonlinearity and  normal dispersion of pulse appear as localized intensity dips on a finite carrier wave background, and they are more robust than bright solitons  \cite{ref3}.  The appearance of multicomponent CNLS type equations as dynamical equations in various areas of physics \cite{ref3,ref4,ref6} and the subsequent studies on these systems lead to the identification of bright, dark, bright-dark, and dark-bright type solitons \cite{ref7,ref8,ref9,ref10,ref11,ref12,ref13,ref14,ref15,ref16}.  Even though there are a number of works on bright and dark soliton propagation and collision separately, still results are scarce for the study on bright-dark type soliton propagation and their collision dynamics. 

In the present paper we perform a study on bright-dark soliton solutions of mixed N-CNLS equations and examine the shape changing collisions of multicomponent bright solitons in the presence of dark components.  We consider the following set of integrable mixed N-CNLS equations (in dimensionless form)
\begin{subequations}            
\bea
iq_{j,z}+q_{j,tt}+2\Big(\sum_{l=1}^{N}\sigma_{l}|q_l|^2 \Big)q_j=0,\qquad j=1, 2, \ldots, N, 
\label{oe}
\eea
where $q_j,\; j=1, 2, \ldots, N$, is the complex amplitude of the $j$th component, the subscripts $z$ and $t$ denote the partial derivatives with respect to normalized distance and retarded time respectively, and the coefficients ${\sigma_l}$'s define the sign of the nonlinearity.  For convenience and without loss of generality, we define ${\sigma_l}$'s for this mixed case as
\bea
\sigma_l&=&1 \quad\quad \mbox{for} \quad l=1, 2, \ldots, m, \nonumber \\
&=&-1 \quad \mbox{for} \quad l=m+1, m+2, \ldots, N.  
\eea
\label{o}
\end{subequations}

Recent theoretical and experimental studies show that the bright solitons in the focusing case ($\sigma_l=1,\quad l=1, 2, \ldots,N $) undergo fascinating shape changing collisions characterized by intensity redistribution, amplitude dependent phase-shift and relative separation distance \cite{ref11,ref12,ref13}, whereas the standard elastic collision of dark solitons occurs in the defocusing case ($\sigma_l=-1,\quad l=1, 2, \ldots,N$ ) \cite{ref8,ref14}.  Later, it has been observed in Ref. \cite{ref16} that bright-bright solitons of mixed CNLS equations (\ref{o}) also undergo shape changing collision but of different nature.  Very recently, it has also been shown that the bright solitons in the mixed CNLS system exhibit periodic energy switching during the shape changing collision process in the presence of linear couplings \cite{ref17}.  Now it is of further interest to examine how the bright solitons are influenced by dark solitons and vice-versa in this mixed CNLS system. The main aim of this paper is to investigate the nature of bright-dark soliton solutions of the mixed CNLS equations (1) and their collisions.

The pioneering works of Makhankov \emph{et al.} \cite{ref18} in the context of Bose-Hubbard model and a few recent works \cite{ref19,ref19a} on left-handed materials (LHMs) in nonlinear optics, suggest that the mixed CNLS system studied in our paper could be of considerable physical significance.  Mixed 2-CNLS system can be obtained as the modified Hubbard model (Lindner-Fedyanin system) in the long-wavelength approximation by taking into account the electron-phonon interaction \cite{ref18}.  Here the two component Bose-condensate is a mixture of two ``gases" with attractive and repulsive interboson fields and the bright-dark soliton solution corresponds to the so called drop-bubble solution.  A straightforward generalization of this mixed 2-CNLS system to an arbitrary number of fields given in Ref. \cite{ref18}.  Exactly this is  the system which we have considered in this paper (Eqs.  (\ref{o})) where the first $m$ components have positive sign of nonlinearity and the remaining components possess negative sign of nonlinearity.  The mixed 2-CNLS equations also arise as governing equations for an electromagnetic field propagation in LHMs with Kerr-type nonlinearity \cite{ref19}.  In fact, this case arises when the fields in the medium experience different types of nonlinearity, leading to $\sigma_1=-\sigma_2=1$ or $\sigma_2=-\sigma_1=1$, corresponding  to a medium of effective positive dielectric permittivity $\epsilon_{eff}$ and effective negative magnetic permeability $\mu_{eff}$ or vice-versa.  This suggests that mixed N-CNLS equations could be a possible generalization of multiple electromagnetic fields propagating in LHMs with suitably chosen effective permittivity and effective permeability.

Equations (\ref{o}) with $N=2$ can also be viewed as governing equation for two fields $q_1$ and $q_2^*$ propagating in the anomalous and normal dispersion regimes, respectively (that is, the self-phase modulation (SPM) coefficients are positive and cross-phase modulation (XPM) coefficients are negative in both the components).  A possible physical realization of such type of nonlinearities is multi-field propagation in a quadratic medium with inefficient phase matching \cite{ref20}.  Another important physical realization of equations (1) arises in the context of Boson-Fermion gas mixtures.  For example, the dynamics of two-species condensates is governed by mixed 2-CNLS equations for suitable choice of intraspecies ($a_{11}, a_{22}$) and interspecies ($a_{12}, a_{21}$) scattering lengths \cite{ref21}.  These two-species condensates offer a wider range of possibilities, the main one being the possibility of having a negative interspecies scattering length.  This possibility has been theoretically explored in the context of Feshbach resonance management and realized experimentally for boson-fermion mixtures \cite{ref21a}.  Thus the interspecies interactions $a_{12}$ and $a_{21}$  can be tuned to be negative (attractive type) and positive (repulsive type) scattering lengths, respectively by Feshbach resonance.  It is  also possible to choose the self interactions as $a_{11}>0$, $a_{22}<0$ \cite{ref21b}.  So studying N-CNLS equations (1) of mixed type will provide a better understanding on the dynamics of multi-species condensates with suitably tuned scattering lengths.  In addition, mixed 2-CNLS equations arise in BECs involving two isotopes of the same element, for example isotopes of rubidium ($\mbox{Rb}^{87}$ and $\mbox{Rb}^{85}$) \cite{ref21c}.  In fact, multicomponent BECs support nonlinear waves which do not exist in single component BECs such as domain-wall solitons, bright-dark solitons, etc.  Thus the mixed CNLS system which we have investigated could be of considerable physical relevance and significance in the context of nonlinear optics and matter waves. 
 
The plan of the paper is as follows.  In Sec. II, we briefly present the Hirota's bilinearization procedure for the mixed N-CNLS equations (1) to obtain exact soliton solutions.  Section III is devoted to obtain exact one and two bright-dark soliton solutions for mixed 2-CNLS, 3-CNLS and N-CNLS equations.  The collision dynamics of bright-dark solitons in mixed CNLS equations is given in Sec. IV, where we have pointed out that if the bright solitons appear in more than one component then they undergo shape changing collisions characterized by intensity redistribution, amplitude dependent phase-shift and change in relative separation distance.  We have also discussed the role of dark solitons on the shape changing collisions of bright solitons and also the effect of bright solitons on the dark soliton collisions.  In Sec. V, we summarize the results of our study.  Asymptotic analysis of mixed 2-CNLS and mixed 3-CNLS equations are given in Appendices A and B, respectively.     
\section{BILINEARIZATION METHOD FOR INTEGRABLE MIXED N-CNLS EQUATIONS}
In this section, we briefly outline the procedure to obtain $m$-bright $-$ $n$-dark soliton solution $(m+n=N)$ of the mixed N-CNLS equations using Hirota's bilinearization method \cite{ref22}.  We denote the soliton solution of Eqs. (1) in which the bright and dark solitons are split up in the first $m$ components and the remaining $(N-m)=n$ components, respectively, as ``mixed soliton solutions", for brevity.  To start with, let us apply the bilinearizing transformation
\begin{subequations}
\bea
q_j&=&\frac{g^{(j)}}{f},\quad \quad j=1, 2, \ldots, m,\\
\label{e1a} 
q_{l+m}&=&\frac{h^{(l)}}{f},\quad \quad l=1, 2, \ldots, n, 
\label{e1b}
\eea
\label{e1}
\end{subequations}
to Eqs. (1), where $g^{(j)}$'s and $h^{(l)}$'s  are arbitrary complex functions of $z$ and $t$ while  $f$ is a real function. Then the set of mixed N-CNLS  equations given by Eqs. (1) reduces to the following set of bilinear equations: 
\begin{subequations}
\bea
&&D_1\left(g^{(j)}\cdot f\right)=0,\qquad j=1, 2, \ldots, m,\\
&&D_1\left(h^{(l)}\cdot f\right)=0,\qquad l=1, 2, \ldots, n,\\
&&D_2(f\cdot f)=2\Bigg(\sum_{j=1}^{m}g^{(j)}g^{(j)*}-\sum_{l=1}^{n}h^{(l)}h^{(l)*}\Bigg),
\label{e2}
\eea
\end{subequations}
where $D_1=(iD_z+D_t^2-\lambda)$, $D_2=(D_t^2-\lambda)$, $*$ denotes complex conjugate and $\lambda$ is a constant to be determined. The Hirota's bilinear operators $D_z$ and $D_t$ are defined as
\bea
D_z^{p}D_t^{q}(a\cdot b) =\bigg(\frac{\partial}{\partial z}-\frac{\partial}{\partial z'}\bigg)^p\bigg(\frac{\partial}{\partial t}-\frac{\partial}{\partial t'}\bigg)^q a(z,t)b(z',t')|_{ \ds (z=z', t=t')}.
\eea
Expanding $g^{(j)}$'s, $h^{(l)}$'s and $f$ formally as power series expansions in terms of a small arbitrary real parameter $\chi$,
\begin{subequations}
\bea
g^{(j)}=&&\chi g_1^{(j)} +\chi^3 g_3^{(j)}+\ldots,\;\;\;\; j=1, 2, \ldots, m,\\
h^{(l)}=&&h_0^{(l)}(1+\chi^2 h_2^{(l)} +\chi^4 h_4^{(l)}+\ldots),\;\;\;\; l=1, 2, \ldots, n,\\
f=&&1+\chi^2 f_2+\chi^4 f_4+\ldots,
\eea
\label{pe}
\end{subequations} \\
and solving the resultant set of equations  recursively, we can obtain the explicit forms of  $g^{(j)}$'s, $h^{(l)}$'s and $f$.

\section{EXACT MIXED TYPE (BRIGHT-DARK) SOLITON SOLUTIONS OF MULTICOMPONENT MIXED CNLS EQUATIONS}
A significant feature of the system of integrable mixed CNLS equations (1) is that it admits a rich structure of soliton solutions like bright, dark, bright-dark soliton type solutions depending upon the boundary conditions.
Here our focus is on bright-dark (or dark-bright) type solutions.  In order to understand the nature of such mixed soliton solutions, their propagation and collision dynamics, we first obtain the solution for the $N=2$ case.  Then by extending the analysis to arbitrary $N$ component case in the remaining sections, we point out that for $N>2$, if the bright solitons are split up in two or more components, there occurs shape changing collision of bright solitons which is also influenced by the presence of dark soliton.
  
\subsection{Bright-dark soliton solutions of the mixed 2-CNLS equations}
\begin{figure}
\centering
\epsfig{figure=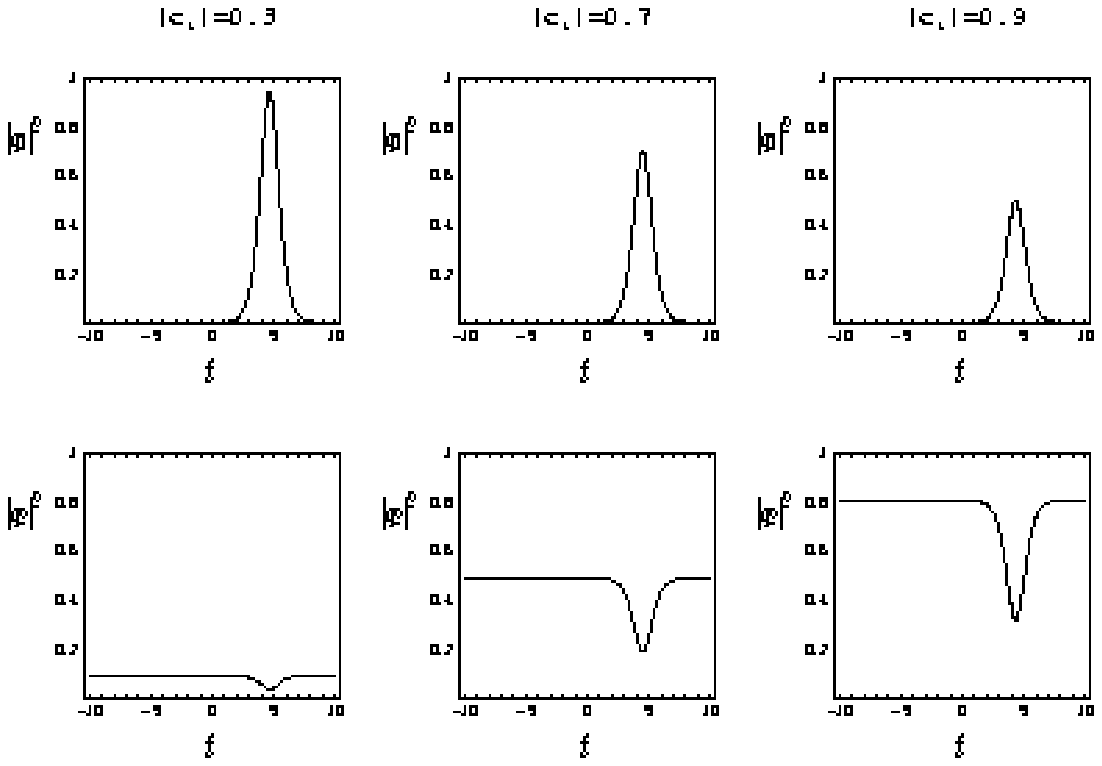, width=0.9\columnwidth}
\caption{Intensity plots of one-soliton of the mixed CNLS equations with $N=2$ for different values of the background parameter $c_1$ for a fixed value of $z$.  Note that the intensity of the bright soliton increases as the depth of the dark soliton decreases.  The other parameters are chosen as $k_1=1+i, \alpha_1^{(1)}=1, b_1=0.2$.}
\label{onet}
\end{figure}

\underline {a) Mixed (Bright-dark) one-soliton solution:}\\
In this case the bright soliton appears in the $q_1$ component and the remaining  component $q_2$ comprises of dark soliton (or vice-versa).  This case corresponds to the choice $\sigma_1=1$ and $\sigma_2=-1$ (or $\sigma_1=-\sigma_2=-1$) for which Eqs. (1) become
\begin{subequations} 
\bea
iq_{1,z}+q_{1,tt}+2\Big(|q_1|^2-|q_2|^2 \Big)q_1=0,\\
iq_{2,z}+q_{2,tt}+2\Big(|q_1|^2-|q_2|^2 \Big)q_2=0.
\eea
\label{2cnls}
\end{subequations}
After restricting the power series expansion (\ref{pe}) as
\begin{subequations} 
\bea
g^{(1)}&=&\chi g_1^{(1)},\\
h^{(1)}&=&h_0^{(1)}(1+\chi^2 h_2^{(1)}),\\
f&=&1+\chi^2 f_2,
\eea
\end{subequations}
and solving the resulting set of linear partial differential equations recursively, one can write down the mixed one-soliton solution explicitly as
\begin{subequations}
\bea
q_1&=&\frac{\alpha_1^{(1)}e^{\eta_1}}{1+e^{{\eta_1+\eta_1^*+R}}}=A_1 k_{1R} e^{i\eta_{1I}} \mbox{sech}\left[\eta_{1R}+\frac {R}{2}\right],\\
q_2&=&\frac{c_1\;e^{i\zeta_1}\left[1-\left(\frac{k_1-ib_1}{k_1^*+ib_1}\right) e^{{\eta_1+\eta_1^*+R}}\right]}{1+e^{{\eta_1+\eta_1^*+R}}},
\eea
\label{2o}
where
\bea
&&e^R=\frac{|{\alpha_1^{(1)}}|^2}{(k_1+k_1^*)^2}\bigg(1-\frac{|c_1|^2}{|k_1-ib_1|^2}\bigg)^{-1},\quad A_1=\left(\frac{\alpha_1^{(1)}}{2 k_{1R}}\right) e^{-\frac{R}{2}}, \\
&&\eta_1=k_1t+i (k_1^2-\lambda) z,\quad k_1=k_{1R}+ik_{1I},\quad \alpha_1^{(1)}=\alpha_{1R}^{(1)}+i\alpha_{1I}^{(1)},\label{suffix} \\
&&\zeta_1=-(b_1^2+\lambda)z+b_1t, \quad \lambda=2|c_1|^2, \quad c_1=c_{1R}+ic_{1I},
\eea
with the condition
\bea
&&|c_1|^2 < |k_1-ib_1|^2.
\label{condn1}
\eea
\end{subequations}
Here $\alpha_1^{(1)}, k_1, \mbox{and}\; c_1 $ are arbitrary complex parameters while $b_1$ is a real parameter.  In  the above equations and in the following sections, the suffixes $R$ and $I$ denote the real and imaginary parts, respectively.  The above one-soliton solution is characterized by seven real parameters $\alpha_{1R}^{(1)},\alpha_{1I}^{(1)}, k_{1R}, k_{1I}, c_{1R}, c_{1I}$ and $b_1$ along with the constraint (\ref{condn1}).  In the context of nonlinear optics the quantity $A_1$ defined through the $\alpha$-parameter can be viewed as the polarization vector of the light pulse/beam and $A_1 k_{1R}$ as the amplitude of the bright soliton.   By defining the quantities $\theta$ and $\phi_1^{(1)}$ as $\theta=\mbox{tan}^{-1}\big({\alpha_{1I}^{(1)}}/{\alpha_{1R}^{(1)}}\big)$  and $\phi_1^{(1)}=\mbox{tan}^{-1}\big((k_{1I}-b_1)/k_{1R}\big)$, respectively, Eqs. (8a) and (8b) can be rewritten as
\begin{subequations}
\bea
&&q_1=\sqrt{k_{1R}^2-|c_1|^2 \mbox{cos}^2{\phi_1^{(1)}}}\;\; e^{i \theta}\mbox{sech}{\bigg[k_{1R}(t-2k_{1I}z)+\frac{R}{2}\bigg]}\nonumber\\&&\qquad \qquad \qquad \qquad \qquad \qquad \times\; e^{ik_{1I}t+i(k_{1R}^2-k_{1I}^2-2|c_1|^2)z},\label{oneq1}\\
&&q_2=-c_1e^{i\zeta_1}\bigg(\mbox{cos}{\phi_1^{(1)}}\;\;\mbox{tanh}{\bigg[k_{1R}(t-2k_{1I}z)+\frac{R}{2}\bigg]}+i\mbox{sin}{\phi_1^{(1)}}\bigg).
\eea
\end{subequations}
Now the condition (\ref{condn1}) becomes $k_{1R}^2 > |c_1|^2 \mbox{cos}^2{\phi_1^{(1)}}$.  It can be inferred from Eq. (\ref{oneq1}) that the intensity of the bright soliton increases with a decrease in the magnitude of the dark soliton parameter $c_1$ as shown in Fig. \ref{onet}. (all the quantities in this and rest of the figures are dimensionless).  This is a consequence of the particular type of cross phase modulation coupling given in Eq. (\ref{2cnls}).  Note that the $\alpha$-parameter influences only the central position of the bright and dark solitons and not the intensities.  However, it can be inferred from Eq. (9a) that the $\alpha$-parameters appear in the complex phase [$e^{i \theta}$] of the amplitude part. \\
\underline {b) Mixed two-soliton solution:}\\
The mixed two-soliton solution can be obtained by terminating the power series expansion (\ref{pe}) as
\begin{subequations}
\bea
g^{(1)}&=&\chi g_1^{(1)}+\chi^3 g_3^{(1)},\\
h^{(1)}&=&h_0^{(1)}(1+\chi^2 h_2^{(1)}+\chi^4 h_4^{(1)}),\\
f&=&1+\chi^2 f_2+\chi^4 f_4.
\eea
\end{subequations}
After solving the resulting bilinear equations recursively, the explicit two-soliton solution is obtained as 
\begin{subequations}
\bea
q_1=&&\frac{1}{D}\Big(\alpha_1^{(1)} e^{\eta_1}+\alpha_2^{(1)} e^{\eta_2}+e^{\eta_1+\eta_1^{*}+\eta_2+\delta_{11}}+e^{\eta_2+\eta_2^{*}+\eta_1+\delta_{21}}\Big), \\
q_2=&&\frac{1}{D} \Big[c_1\;e^{i\zeta_1}\Big(1+e^{\eta_1+\eta_1^{*}+Q_{11}^{(1)}}+e^{\eta_1+\eta_2^{*}+Q_{12}^{(1)}}+e^{\eta_2+\eta_1^{*}+Q_{21}^{(1)}}+e^{\eta_2+\eta_2^{*}+Q_{22}^{(1)}}\nonumber\\&&+e^{\eta_1+\eta_1^{*}+\eta_2+\eta_2^{*}+Q_3^{(1)}} \Big)\Big], 
\eea
where
\bea
D=&&1+e^{\eta_1+\eta_1^{*}+R_1}+e^{\eta_1+\eta_2^{*}+\delta_0}+e^{\eta_2+\eta_1^{*}+\delta_0^{*}}+e^{\eta_2+\eta_2^{*}+R_2}\nonumber\\&&+e^{\eta_1+\eta_1^{*}+\eta_2+\eta_2^{*}+R_3}
\label{d}
\eea
and 
\bea
\eta_j=k_jt+i (k_j^2-2|c_1|^2) z, \quad j=1,2.
\eea
\label{2-2}
\end{subequations}
In the above $\alpha_1^{(1)},\alpha_2^{(1)}, k_1, k_2\; \mbox{and}\; c_1$ are complex parameters and $b_1$ is a real parameter (see below).  Introducing the quantity
\begin{subequations} 
\bea
\mu_{ \ds{ij}}=\frac{\alpha_i^{(1)}{\alpha_j^{(1)}}^*}{(k_i+k_j^*)^2}\left[1-\frac{|c_1|^2}{(k_i-ib_1)(k_j^*+ib_1)}\right]^{-1},\quad i,j=1,2,
\eea
the various other parameters in the expression (\ref{2-2}) are defined as follows:
\bea 
&&e^{R_1}=\mu_{11},\quad e^{R_2}=\mu_{22},\quad e^{\delta_0}=\mu_{12},\quad e^{\delta_0^{*}}=\mu_{21},\\
\label{2-er1}
&&e^{\delta_{11}}=\frac{(k_2-k_1)^2}{\alpha_1^{(1)*}}\left[1+\frac{|c_1|^2}{(k_1-ib_1)(k_2-ib_1)}\right]\mu_{11}\mu_{21}, \\
&&e^{\delta_{21}}=\frac{(k_2-k_1)^2}{\alpha_2^{(1)*}}\left[1+\frac{|c_1|^2}{(k_1-ib_1)(k_2-ib_1)}\right]\mu_{22}\mu_{12}, \\
&&e^{Q_{ij}^{(1)}}=-{ \ds \frac{(k_i-ib_1)}{(k_j^*+ib_1)}\mu_{ij}},\quad i,j=1,2, \quad
e^{Q_3^{(1)}}={ \ds\Bigg[ \frac{(k_1-ib_1)(k_2-ib_1)}{(k_1^*+ib_1)(k_2^*+ib_1)}\Bigg]e^{R_3}},\\
\label{q11s} 
\label{Qs}
&&\hspace{-2cm}\mbox{and}\nonumber\\
&&e^{R_3}=\left(\frac{\mu_{11}\mu_{12}\mu_{21}\mu_{22}}{|\alpha_1^{(1)}\alpha_2^{(1)}|^2}\right)|k_1-k_2|^2\bigg|1+\frac{|c_1|^2}{(k_1-ib_1)(k_2-ib_1)}\bigg|^2.
\eea
\end{subequations}
The above two-soliton solution (\ref{2-2}) is restricted by the conditions $k_{jR}^2+(k_{jI}-b_1)^2 > |c_1|^2, j=1,2,$ as in the case of one-soliton solution.  The two-soliton solution (\ref{2-2}) is characterized by eleven real parameters $\alpha_{1R}^{(1)}, \alpha_{2R}^{(1)}, \alpha_{1I}^{(1)}, \alpha_{2I}^{(1)}, k_{1R}, k_{1I},k_{2R}, k_{2I}, c_{1R}, c_{1I} $ and $b_1$ . 
\subsection{Bright-dark soliton solutions of the mixed 3-CNLS equations}
Let us consider the set of mixed 3-CNLS equations which corresponds to Eq. (\ref{o}) with $N=3$.  In its explicit form the set of mixed 3-CNLS equations reads as 
\begin{subequations} 
\bea
iq_{1,z}+q_{1,tt}+2\Big(\sigma_1|q_1|^2+\sigma_2|q_2|^2+\sigma_3|q_3|^2 \Big)q_1=0,\\
iq_{2,z}+q_{2,tt}+2\Big(\sigma_1|q_1|^2+\sigma_2|q_2|^2+\sigma_3|q_3|^2 \Big)q_2=0,\\
iq_{3,z}+q_{3,tt}+2\Big(\sigma_1|q_1|^2+\sigma_2|q_2|^2+\sigma_3|q_3|^2 \Big)q_3=0.
\eea
\end{subequations}
In the above, the nonlinearity coefficients $\sigma_j$'s, $j=1,2,3$ take the values either $\sigma_1=\sigma_2=+1,\; \mbox{and}\;  \sigma_3=-1$ or $\sigma_1=+1\; \mbox{and}\; \sigma_2=\sigma_3=-1$.  In Eqns. (13), in the context of BECs the components $q_1$, $q_2$ and $q_3$ either denote the condensates of three isotopes of the same element (for example Rb isotopes) \cite{ref21c} or the hyperfine spin states of spinor BECs \cite{ref22b}.  This set of equations admits the following two distinct types of mixed soliton solutions.\\
\underline{(i) 2-bright $-$ 1-dark soliton solution:}\\
In this case the bright solitons are separated out into two of the three components and the dark soliton appears in the remaining component (with the choice, $\sigma_1=\sigma_2=1,\; \mbox{and}\;  \sigma_3=-1$).\\
\underline{(ii) 1-bright  $-$ 2-dark soliton solution:}\\
In this type of solution the dark solitons appear in two components and the remaining component comprises of the bright soliton (with the choice, $\sigma_1=1\; \mbox{and}\; \sigma_2=\sigma_3=-1$).\\
The procedure of obtaining these soliton solutions is similar to that of the mixed 2-CNLS equations. 
\subsubsection{2-bright $-$ 1-dark soliton solution}
\begin{figure}
\centering
\epsfig{figure=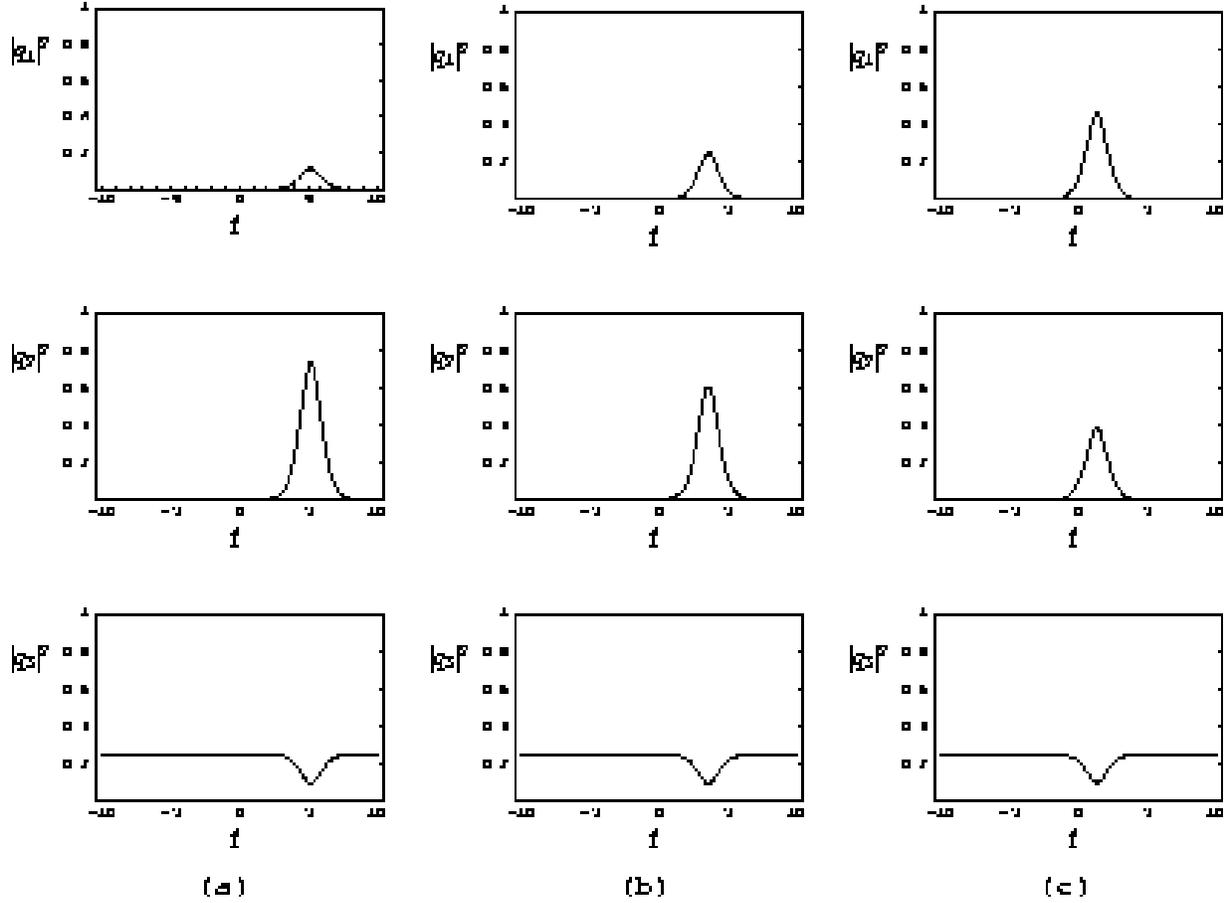, width=1.0\columnwidth}
\caption{Intensity plots of one-soliton of the mixed CNLS equations with $N=3$ for different values of the $\alpha $ parameters for a fixed value of $z$. (a) $\alpha_1^{(1)}=0.2+0.01 i, \alpha_1^{(2)}=0.5+0.05 i$, (b)  $\alpha_1^{(1)}=1+i, \alpha_1^{(2)}=2+i$ and (c)  $\alpha_1^{(1)}=13-13 i, \alpha_1^{(2)}=17+0.3 i$.  The role of $\alpha$ parameters are seen both in the intensity and phase of the bright soliton while it affects the phase (central position) of the dark soliton.  The parameters $c_1, b_1$, and $k_1$ are chosen as $|c_1|=0.56, b_1=0.2, k_1=1+i$.}
\label{onea}
\end{figure}

We present below the explicit forms of the obtained one- and two-soliton solutions.\\
\underline {a) Mixed one-soliton solution:}\\
The mixed one-soliton solution, in which two bright solitons appear in the first two components and the dark one appears in third component, is obtained by Hirota's method as 
\begin{subequations}
\bea
q_j=&&\frac{\alpha_1^{(j)}e^{\eta_1}}{1+e^{{\eta_1+\eta_1^*+R}}},\quad j=1,2, \\
=&&A_j k_{1R} e^{i\eta_{1I}} \mbox{sech}\Big[\eta_{1R}+\frac {R}{2}\Big],\\
q_3=&&\frac{c_1\;e^{i\zeta_1}\left[1-\left(\frac{k_1-ib_1}{k_1^*+ib_1}\right) e^{{\eta_1+\eta_1^*+R}}\right]}{1+e^{{\eta_1+\eta_1^*+R}}},
\eea
where 
\bea
e^R=&&\frac{\sum_{j=1}^{2}(\alpha_1^{(j)}\alpha_1^{(j)*})}{(k_1+k_1^*)^2}\bigg(1-\frac{|c_1|^2}{|k_1-ib_1|^2}\bigg)^{-1},\\
A_j=&&\left(\frac{\alpha_1^{(j)}}{2 k_{1R}}\right) e^{-\frac{R}{2}}\equiv \frac{\sqrt{k_{1R}^2-|c_1|^2 \mbox{cos}^2{\phi_1^{(1)}}}}{k_{1R}}\left(\frac{\alpha_1^{(j)}}{\sqrt{(|\alpha_1^{(1)}|^2+|\alpha_1^{(2)}|^2)}}\right) ,\quad j=1,2.
\eea
\end{subequations}
Here the quantities $\eta_1$ and $\zeta_1$ are as defined in Eq. (8d) and (8e),  respectively.  $A_j$'s defined through the $\alpha^j$ and $c_1$ parameters represent the polarization of the bright components.  For the dark component, the parameter $b_1$ denotes the direction of the background and $c_1$ gives its amplitude.  The one-soliton solution is characterized by nine real parameters $\alpha_{1R}^{(1)}, \alpha_{1I}^{(1)}, \alpha_{1R}^{(2)}, \alpha_{1I}^{(2)}, k_{1R}, k_{1I}, c_{1R}, c_{1I} $ , and $b_1$ and is restricted by the condition $|c_1|^2 < |k_1-ib_1|^2$.  Now the role of $\alpha$-parameters can be realized explicitly in the amplitude (intensity) of bright components and also through the non-trivial phase of all the components.  This is shown in Fig. \ref{onea}.  In fact this has important consequences in the collision process as will be illustrated in the following sections.  Thus the dark soliton part influences the bright part through the parameters $c_1$ and $b_1$ whereas the bright solitons influence the dark soliton phase (central position) through the $\alpha$ - parameters.\\
\underline {b) Two-soliton solution:}\\
Following the Hirota's bilinearization method as in the case of $N=2$, here we obtain the two-soliton solution as 
\begin{subequations}
\bea
q_j=&&\frac{1}{D}\Big(\alpha_1^{(j)} e^{\eta_1}+\alpha_2^{(j)} e^{\eta_2}+e^{\eta_1+\eta_1^{*}+\eta_2+\delta_{1j}}+e^{\eta_2+\eta_2^{*}+\eta_1+\delta_{2j}}\Big),\quad j=1,2,\\
q_3=&&\frac{1}{D} \Big[c_1\;e^{i\zeta_1}\Big(1+e^{\eta_1+\eta_1^{*}+Q_{11}^{(1)}}+e^{\eta_1+\eta_2^{*}+Q_{12}^{(1)}}+e^{\eta_2+\eta_1^{*}+Q_{21}^{(1)}}+e^{\eta_2+\eta_2^{*}+Q_{22}^{(1)}}\nonumber\\&&+e^{\eta_1+\eta_1^{*}+\eta_2+\eta_2^{*}+Q_3^{(1)}} \Big)\Big],\\
e^{\delta_{1j}}=&&(k_2-k_1)\mu_{11}\mu_{21}(\alpha_2^{(j)}\chi_{21}-\alpha_1^{(j)}\chi_{11}),\\
e^{\delta_{2j}}=&&(k_2-k_1)\mu_{12}\mu_{22}(\alpha_2^{(j)}\chi_{22}-\alpha_1^{(j)}\chi_{12}),\\
e^{R_3}=&&|k_1-k_2|^2\mu_{11}\mu_{12}\mu_{21}\mu_{22}{\ds(\chi_{12}\chi_{21}-\chi_{11}\chi_{22})}, \quad j=1,2,
\eea
\label{es}
where $\mu_{il}$'s are now redefined as
\bea
&&\mu_{il}=\frac{1}{(k_i+k_l^*)\chi_{il}},\\
&&\chi_{il}=\frac{(k_i+k_l^*)}{\sum_{j=1}^{2}(\alpha_i^{(j)}\alpha_l^{(j)*})}\bigg(1-\frac{|c_1|^2}{(k_i-ib_1)(k_l^*+ib_1)}\bigg),\quad i,l=1,2. 
\eea
\end{subequations}
The form of $D$ is given as in Eq. (\ref{d}) and the expressions for $e^{Q_{11}^{(1)}}$, $e^{Q_{12}^{(1)}}$, $e^{Q_{21}^{(1)}}$, $e^{Q_{22}^{(1)}}$ and $e^{Q_{3}^{(1)}}$ take the form as given in  Eq. (12e) with the above redefinition of the  $\mu_{il}$'s.  $\alpha_{iR}^{(j)}, \alpha_{iI}^{(j)}, k_{iR}, k_{iI},i,j=1,2,c_{1R}, c_{1I} $ and $b_1$ are the fifteen real parameters which characterize the above solution.  The nature of this two-soliton solution will be discussed in Sec. IVB.
\subsubsection{1-bright $-$ 2-dark soliton solution }
Next we consider the case where the bright soliton appears in the $q_1$ component   and the two dark solitons are found in the remaining two components $(q_2, q_3)$.  This gives us the possibility of introducing two background fields $c_1 e^{i \zeta_1}$ and $c_2 e^{i \zeta_2}$.\\ 
\underline {a) Mixed one-soliton solution:}\\
The corresponding one-soliton solution obtained by using Hirota's method is 
\begin{subequations}
\bea
q_1=&&\frac{\alpha_1^{(1)} e^{\eta_1}}{1+e^{{\eta_1+\eta_1^*+R}}} , \\
=&&A_1 k_{1R} e^{i\eta_{1I}} \mbox{sech}\left[\eta_{1R}+\frac {R}{2}\right],\\
q_{l+1}=&&\frac{c_l\;e^{i\zeta_l}\left[1-{\ds\left(\frac{k_1-ib_l}{k_1^*+ib_l}\right)} e^{{\eta_1+\eta_1^*+R}}\right]}{1+e^{{\eta_1+\eta_1^*+R}}}, \quad l=1,2,
\eea
where
\bea
&&\eta_1=k_1t+i (k_1^2-\lambda) z,\quad \zeta_l=-(b_l^2+\lambda)z+b_lt, \quad \lambda=2(|c_1|^2+|c_2|^2), \quad l=1,2.
\nonumber\\
&&e^R=\frac{|{\alpha_1^{(1)}}|^2}{(k_1+k_1^*)^2}\bigg[1-\frac{|c_1|^2}{|k_1-ib_1|^2}-\frac{|c_2|^2}{|k_1-ib_2|^2}\bigg]^{-1},\\
&&A_1=\left(\frac{\alpha_1^{(1)}}{2 k_{1R}}\right) e^{-\frac{R}{2}}.
\eea
\end{subequations}
Now the one-soliton solution is characterized by ten real parameters, $ \alpha_{1R}^{(1)}, \alpha_{1I}^{(1)}$,  $k_{1R}, k_{1I}$, $c_{1R}, c_{1I}$, $c_{2R}, c_{2I}$, $b_1$ and $b_2 $ .\\
\underline {b) Two-soliton solution:}\\
As in the previous section, here also we obtain the two-soliton solution as
\begin{subequations}
\bea
q_1=&&\frac{1}{D}\Big(\alpha_1^{(1)} e^{\eta_1}+\alpha_2^{(1)} e^{\eta_2}+e^{\eta_1+\eta_1^{*}+\eta_2+\delta_{11}}+e^{\eta_2+\eta_2^{*}+\eta_1+\delta_{21}}\Big), \\
q_{l+1}=&&\frac{1}{D} \Big[c_l\;e^{i\zeta_l}\Big(1+e^{\eta_1+\eta_1^{*}+Q_{11}^{(l)}}+e^{\eta_1+\eta_2^{*}+Q_{12}^{(l)}}+e^{\eta_2+\eta_1^{*}+Q_{21}^{(l)}}\nonumber\\&&+e^{\eta_2+\eta_2^{*}+Q_{22}^{(l)}}+e^{\eta_1+\eta_1^{*}+\eta_2+\eta_2^{*}+Q_3^{(l)}} \Big)\Big],\quad l=1,2,
\eea
where
\bea
&&e^{\delta_{11}}=(k_2-k_1)^2\frac{\mu_{11}\mu_{21}}{{\alpha_1^{(1)}}^*}\rho,\quad e^{\delta_{21}}=(k_2-k_1)^2\frac{\mu_{12}\mu_{22}}{{\alpha_2^{(1)}}^*}\rho.
\eea
In Eqs. (17a) and (17b), the form of $D$, $e^{Q_{ij}^{(l)}}$ and $e^{Q_3^{(l)}}$ are same as in Eqs. (11c) and (12e) with $i,j,l=1,2$.  Also, the quantities $e^{R_3}$, $\mu_{il}$ and $\rho$ are redefined as 
\bea
&&e^{R_3}=|k_1-k_2|^2\frac{\mu_{11}\mu_{12}\mu_{21}\mu_{22}}{|\alpha_1^{(1)} \alpha_2^{(1)}|^2}|\rho|^2,\\
&&\mu_{ \ds{il}}=\frac{\alpha_i^{(1)}{\alpha_l^{(1)}}^*}{(k_i+k_l^*)^2}\left[1-\sum_{v=1}^2\frac{|c_v|^2}{(k_i-ib_v)(k_l^*+ib_v)}\right]^{-1},\quad i,l=1,2\\
&&\hspace{-3cm}\mbox{and}\nonumber\\
&&\rho=\left[1+\sum_{v=1}^2\frac{|c_v|^2}{(k_1-ib_v)(k_2-ib_v)}\right].
\eea
\end{subequations}
The two-soliton solution is characterized by fourteen real parameters.  Again we will study the nature of this solution in Sec. IVB.
\subsection{N-soliton solutions}
The above procedure of obtaining soliton solutions can be extended to three- and N-soliton solutions with some effort, though the analysis is cumbersome.  In this work,  we restrict our analysis to the two soliton solution only as the N-soliton collisions represented by N-soliton solution in general take place pair-wise in soliton theory.  Work is in progress in this direction and the results will be published separately.  
\subsection{Bright-dark soliton solutions of mixed N-CNLS case}
After obtaining the two and three component mixed soliton solutions, the next natural step is to generalize the results to arbitrary N component case, where $N=m+n$.  For this purpose, we consider the case where the bright solitons appear in the first m components and the dark solitons appear in the remaining $n\; (\equiv (N-m))$ components.  So the resulting mixed soliton solution can be denoted as $m$-bright$-$$n$-dark type soliton solution, as pointed out in Sec. II. 
\subsubsection{m-bright$-$n-dark soliton solution}
\underline {a) One-soliton solution:}\\ 
The mixed one-soliton solution of the mixed N-CNLS case is found as
\begin{subequations}
\bea
q_j=&&\frac{\alpha_1^{(j)} e^{\eta_1}}{1+e^{{\eta_1+\eta_1^*+R}}} , \\
=&&A_j k_{1R} e^{i\eta_{1I}} \mbox{sech}\Big[\eta_{1R}+\frac {R}{2}\Big],\quad j=1, 2, \ldots, m,\\
q_{l+m}=&&\frac{c_l\;e^{i\zeta_l}\left[1-\left(\frac{k_1-ib_l}{k_1^*+ib_l}\right) e^{{\eta_1+\eta_1^*+R}}\right]}{1+e^{{\eta_1+\eta_1^*+R}}}, \quad l=1, 2, \ldots, n,
\eea
where
\bea
e^R=&&\frac{\sum_{j=1}^{m}(\alpha_1^{(j)}\alpha_1^{(j)*})}{(k_1+k_1^*)^2}\bigg(1-\sum_{l=1}^{n}\frac{|c_l|^2}{|k_1-ib_l|^2}\bigg)^{-1},\\
A_j=&&\left(\frac{\alpha_1^{(j)}}{2 k_{1R}}\right) e^{-\frac{R}{2}}.
\eea
Here
\bea
\eta_j&=&k_jt+i (k_j^2-\lambda) z,\quad \zeta_l=-(b_l^2+\lambda)z+b_lt, \quad\lambda=2 \sum_{l=1}^n |c_l|^2,
\\
j&=&1, 2, \ldots, m, \mbox{and} \; l=1, 2, \ldots, n.\nonumber
\eea
\end{subequations}
The one-soliton solution is characterized by $(2m+3n+2)$ number of real parameters, $\alpha_{1R}^{(j)}, \alpha_{1I}^{(j)},  k_{1R}, k_{1I}, c_{lR}, c_{lI}, b_l,\quad j=1, 2, \ldots, m,\quad l=1, 2, \ldots, n$ with the condition $|c_l|^2 < |k_1-ib_l|^2, l=1, 2, \ldots, n$.  \\
\underline {b)Two-soliton solution:}\\
Generalization of the mixed two-soliton solution presented in the previous subsections for $N=2$ and $N=3$ cases yields the following m-bright$-$n-dark two-soliton solution of Eqs. (1) with arbitrary $N$:
\begin{subequations}
\bea
&&q_j=\frac{1}{D}\Big(\alpha_1^{(j)}e^{\eta_1}+\alpha_2^{(j)} e^{\eta_2}+e^{\eta_1+\eta_1^{*}+\eta_2+\delta_{1j}}+e^{\eta_2+\eta_2^{*}+\eta_1+\delta_{2j}}\Big),\nonumber\\&& \qquad \qquad \qquad \qquad \qquad \qquad \qquad \qquad  j=1,2, \ldots,m,\\
&&q_{l+m}=\frac{1}{D} \Big[c_l\;e^{i\zeta_l}\Big(1+e^{\eta_1+\eta_1^{*}+Q_{11}^{(l)}}+e^{\eta_1+\eta_2^{*}+Q_{12}^{(l)}}+e^{\eta_2+\eta_1^{*}+Q_{21}^{(l)}}\nonumber\\&&\qquad+e^{\eta_2+\eta_2^{*}+Q_{22}^{(l)}}+e^{\eta_1+\eta_1^{*}+\eta_2+\eta_2^{*}+Q_3^{(l)}} \Big)\Big],\qquad l=1,2, \ldots,n.
\eea 
Here the denominator $D$ is given by Eq. (\ref{d}).  The quantities $ e^{R_1}, e^{R_2}, e^{\delta_0}$, and $e^{\delta_0^*}$ are as defined in Eq. (12b) but with the following redefinitions 
\bea
&&e^{Q_{ij}^{(l)}}=-{ \ds \frac{(k_i-ib_l)}{(k_j^*+ib_l)}\mu_{ij}}, \quad i,j=1,2, \quad
e^{Q_3^{(l)}}={ \ds \Bigg[\frac{(k_1-ib_l)(k_2-ib_l)}{(k_1^*+ib_l)(k_2^*+ib_l)}\Bigg]e^{R_3}},\\
&&e^{\delta_{1j}}=(k_2-k_1)\mu_{11}\mu_{21}(\alpha_2^{(j)}\chi_{21}-\alpha_1^{(j)}\chi_{11}),\quad j=1,2, \ldots,m,\nonumber\\
&&e^{\delta_{2j}}=(k_2-k_1)\mu_{12}\mu_{22}(\alpha_2^{(j)}\chi_{22}-\alpha_1^{(j)}\chi_{12})\nonumber,\\
&&e^{R_3}=|k_1-k_2|^2\mu_{11}\mu_{12}\mu_{21}\mu_{22}{\ds[\chi_{12}\chi_{21}-\chi_{11}\chi_{22}]},\\
&&\mu_{ip}=\frac{1}{(k_i+k_p^*)\chi_{ip}}\nonumber\\
&&\hspace{-2cm}\mbox{and}\nonumber\\
&&\chi_{ip}=\frac{(k_i+k_p^*)}{\sum_{j=1}^{m}(\alpha_i^{(j)}\alpha_p^{(j)*})}\bigg(1-\sum_{l=1}^{n}\frac{|c_l|^2}{(k_i-ib_{l})(k_p^*+ib_{l})}\bigg),\quad i,p=1,2.
\eea
\end{subequations}
The number of real parameters which characterize the two-soliton solution is $(4m+3n+4)$.  
\section{SHAPE CHANGING COLLISIONS OF MIXED SOLITONS}
The fascinating property of the bright solitons of the integrable N-CNLS system with focusing type nonlinearity is that they exhibit shape changing collisions characterized by intensity redistribution, amplitude dependent phase-shift and relative separation distances \cite{ref12,ref13,ref16}, which can then be used to construct collision based logic gates for optical computation \cite{ref23,ref24}.  In this and in the following sections we analyze such collision dynamics of bright solitons in the presence of dark solitons in the mixed 2-CNLS and 3-CNLS equations and also its effect on the propagation and collision of dark solitons.  In this regard, we perform an asymptotic analysis of the two-soliton solutions for mixed 2-CNLS and 3-CNLS equations.
\subsection{Asymptotic analysis of mixed two-soliton solution of mixed 2-CNLS equations  }
To start with, we consider the collision properties associated with the mixed two-soliton solution (11) of the mixed 2-CNLS equations (\ref{2cnls}).  Following this we carry out the analysis for mixed 3-CNLS equations.  Without loss of generality, we take $k_{jR}>0$ and $k_{1I}>k_{2I}$, $k_j=k_{jR}+ik_{jI}, j=1, 2$,  and obtain the asymptotic forms of two colliding solitons (say $S_1$ and $S_2$).  Similar analysis can be carried out for other choices of $k_{jR}$ and $k_{jI}$ also.  Using the expression (11) for the bright-dark two soliton solution of the mixed 2-CNLS system (\ref{2cnls}), we carry out a detailed asymptotic analysis in Appendix A for the two soliton collision process.  Based on this analysis we identify the following:

\subsubsection{ The role of dark soliton on bright soliton collision:}
\begin{figure}
\centering
\epsfig{figure=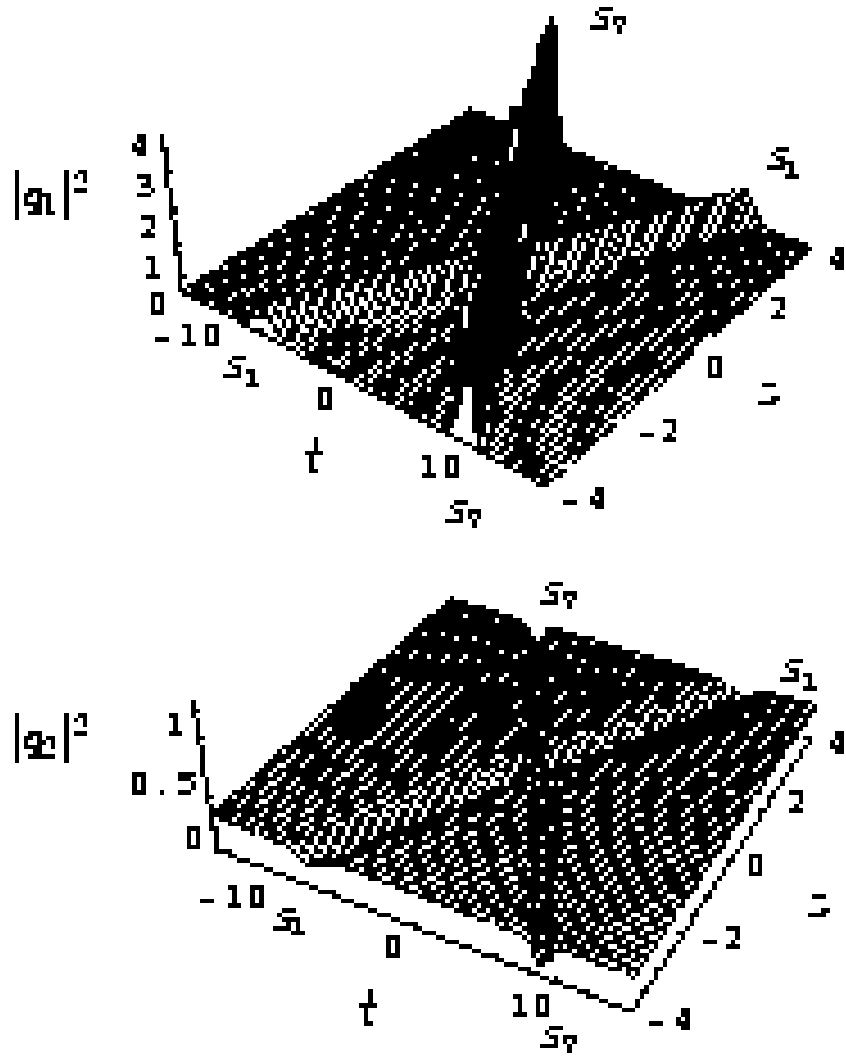, width=0.7\columnwidth}
\caption{Elastic collision of (bright-dark) two-solitons  in the mixed CNLS system for the $N=2$ case.  The parameters are chosen as given in the text.}
\label{1b-1d}
\end{figure}

\begin{figure}
\centering
\epsfig{figure=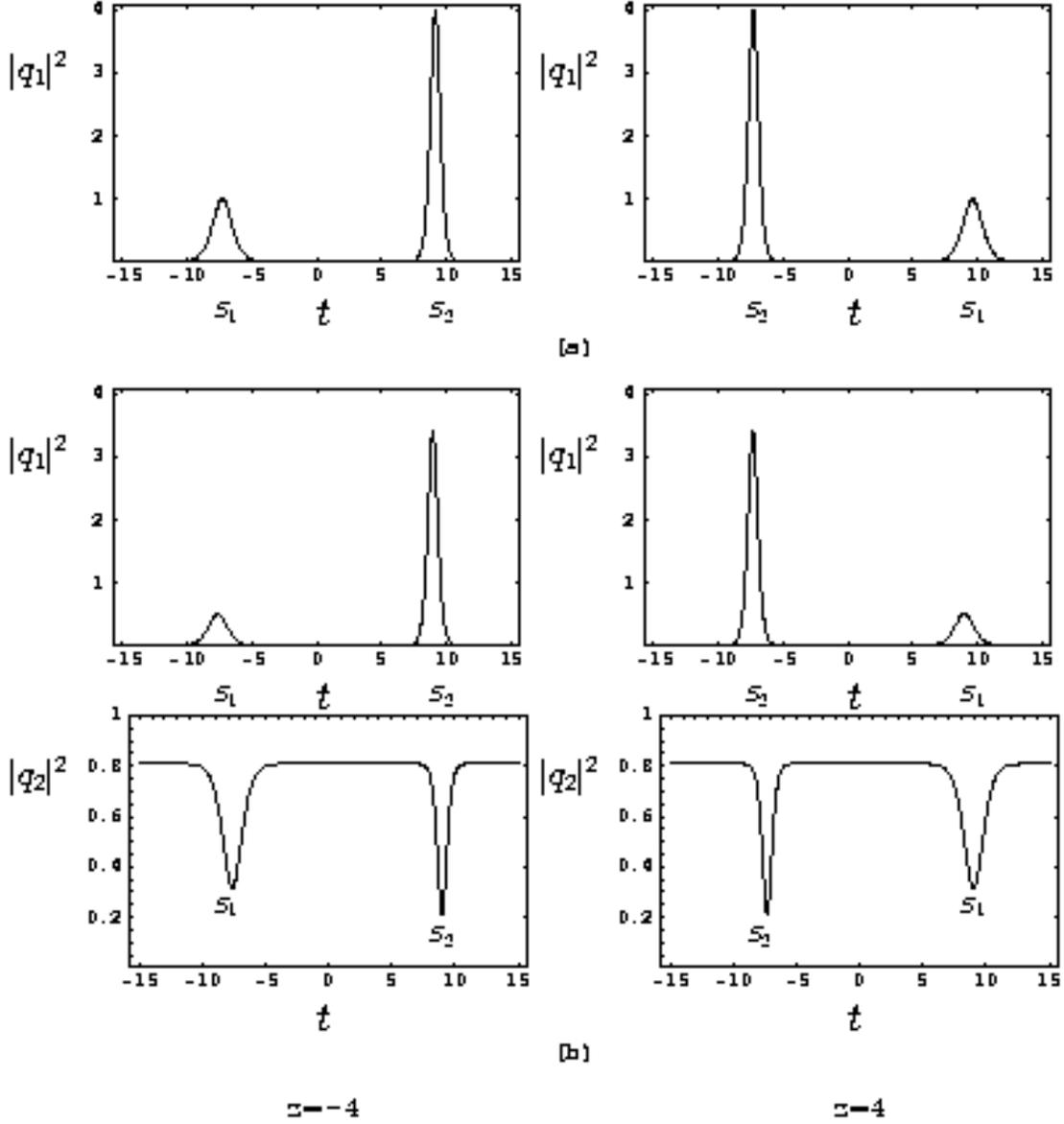, width=0.9\columnwidth}
\caption{Intensity profiles of two colliding bright solitons of the mixed CNLS equations with $N=2$, before (z=-4) and after (z=4) collision: (a) in the absence of dark component ($c_1=0$ in Eq. (11)); (b) in the presence of dark component  ($c_1\neq 0$ in Eq. (11)).  The figure is plotted for special choice of parameters (as given in the text) with $|c_1|=0.56$.  Note the elastic nature of the collision.}
\label{two}
\end{figure}

The amplitudes of the two solitons $S_1$ and $S_2$ before (after) interaction are given by $A_1^{1-} k_{1R}\; (A_1^{1+} k_{1R}) $ and $A_1^{2-}k_{2R}\; (A_1^{2+} k_{2R}) $, respectively, in the $q_1$ component.  Forms of $A_1^{j\pm}, j=1,2,$ are given in Appendix A (see Eqs. (A1c), (A2c), (A3c) and (A4c)).  By rewriting these forms  one can show that the intensities of solitons before and after interaction are same (elastic), (i.e.) $|A_1^{j-}|=|A_1^{j+}|,\; j=1,2$, even though the complex amplitudes differ in phase.  Also, the appearance of the background parameter `$c_1$' in the expression for the bright soliton amplitudes before and after collision (see Eqs. (A1c), (A2c), (A3c) and (A4c)]) shows that this parameter influences the bright soliton amplitudes throughout the collision.\\
Typical bright-dark soliton collision in the mixed 2-CNLS system is shown in Fig. \ref{1b-1d} for the parametric choices $k_1=1+i, k_2=2-i, |c_1|=0.56, b_1=0.2, \alpha_1^{(1)}=1, \alpha_2^{(1)}=1+i$.  For better understanding, first we plot the intensity profiles showing the bright soliton elastic collision scenario in the absence of dark soliton parameter ($c_1=0$), at $z=-4$ and $z=4$ .  This is shown in Fig. \ref{two}(a).  Then the same bright soliton collision is plotted for the above parametric choices as in Fig. \ref{1b-1d} at $z=-4$ and $z=4$, that is in the presence of the dark component, in Fig. \ref{two}(b).  From these two figures, we observe that due to the presence of dark soliton the amplitudes of the colliding bright solitons are reduced by the same amount throughout the collision.\\
The two colliding solitons $S_1$ and $S_2$ suffer phase-shifts $\Phi_1$ and $\Phi_2$, respectively in both the bright and dark components.  These phase-shifts for the bright and dark solitons are given by the expression 
\begin{subequations}
\bea
\Phi_1=-\Phi_2&=&\bigg( \frac{R_3-R_2-R_1}{2}\bigg),\nonumber \\
&=&\frac{1}{2}\log\Bigg(\frac{|k_1-k_2|^2}{(k_1+k_2^*)^2(k_2+k_1^*)^2}\frac{{\big[1+\frac{|c_{1}|^2}{P_1P_2}\big]}{\big[1+\frac{|c_{1}|^2}{P_1^*P_2^*}\big]}}{{\big[1-\frac{|c_{1}|^2}{P_2P_1^*}\big]}{\big[1-\frac{|c_{1}|^2}{P_1P_2^*}\big]}}\Bigg),
\eea
where $P_1=k_1-ib_1$ and $P_2=k_2-ib_1$.  Note that the phase-shifts appearing here (Eq. (20a)) and in the following (see Eqs. (23a) and (26)) are real quantities as the terms appearing in the argument of log function are products of complex conjugates.  The role of dark component comes into picture through this phase-shift due to the explicit appearance of the background parameters $(c_1, b_1)$.  Notice that in the absence of the dark component ($c_1=0$), the phase-shift $\Phi_1$ reduces to the standard phase-shift experienced by colliding solitons in scalar nonlinear Schr{\"o}dinger equations (NLS).  Also, it is important to notice that the $\alpha$-parameters have no effect on the phase-shift in the mixed 2-CNLS case.  However, this is not true in the case of mixed N-CNLS case for more than two components ($N>2$) as will be shown in the following sections.  The relative separation distances between the two colliding solitons before interaction
\bea
t_{12}^{-}=\frac{R_3 k_{1R}-R_1 (k_{1R}+k_{2R})}{2k_{1R}k_{2R}}
\eea
and after interaction
\bea
t_{12}^{+}=\frac{R_2(k_{1R}+k_{2R})-R_3 k_{2R}}{2k_{1R}k_{2R}}
\eea
\end{subequations}
also remain unaffected by $\alpha$-parameters.
\subsubsection{The role of bright soliton on dark soliton collision:}
The asymptotic expressions [Eqs. (A1b), (A2b), (A3b) and (A4b)] reveal the fact that the collision among dark solitons is elastic as the intensities of colliding solitons remain the same before and after interaction.  From Fig. \ref{1b-1d} it can be observed that the bright component parameter `$\alpha$' has no influence at all either in the amplitude or in the phase-shift of the dark soliton during collision.  Thus the inclusion of $\alpha$-parameters in the bright soliton solution (see Eq. (11)) does not affect dark soliton collisions. 
 
\subsection{Asymptotic analysis of mixed two-soliton solution of mixed 3-CNLS equations} 
The next natural step is to study the collision process in the mixed 3-CNLS equations and one can generalize the results to the N-CNLS equations with arbitrary $N$.  The asymptotic expressions of the solitons corresponding to the $N=3$ case are presented in Appendix B.  This 3-CNLS system admits two distinct types of solution as mentioned in Sec. IIIB.  First let us consider the 2-bright$-$1-dark soliton collision and the 1-bright$-$2-dark soliton collision.  
\subsubsection{2-Bright$-$1-Dark Soliton Collision:}
\begin{figure}
\centering
\epsfig{figure=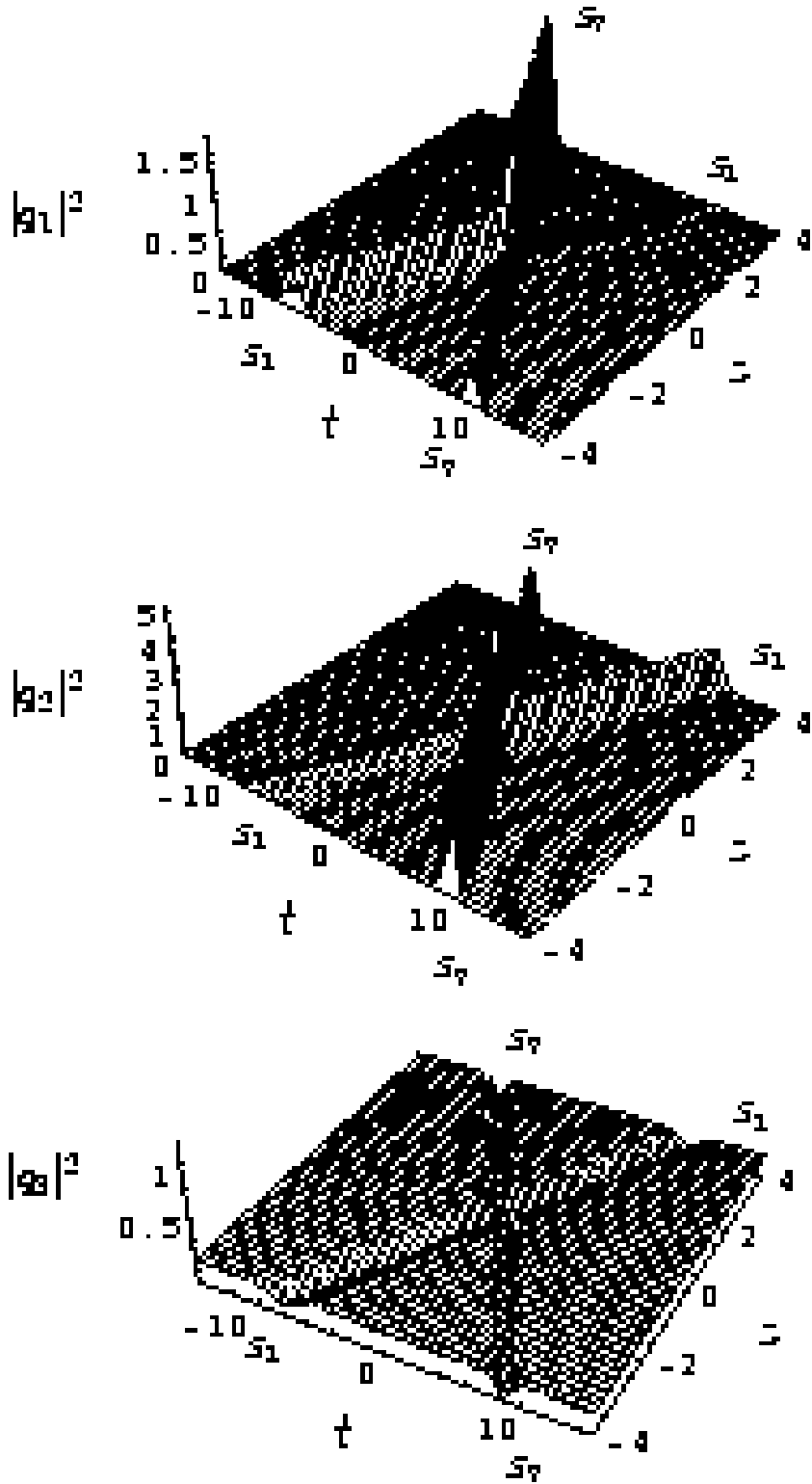, width=0.7\columnwidth}
\caption{Shape changing collision of two-solitons  in the mixed CNLS system for the $N=3$ case.  The parameters are as given in the text.}
\label{2b-1d}
\end{figure}

\begin{figure}
\centering
\epsfig{figure=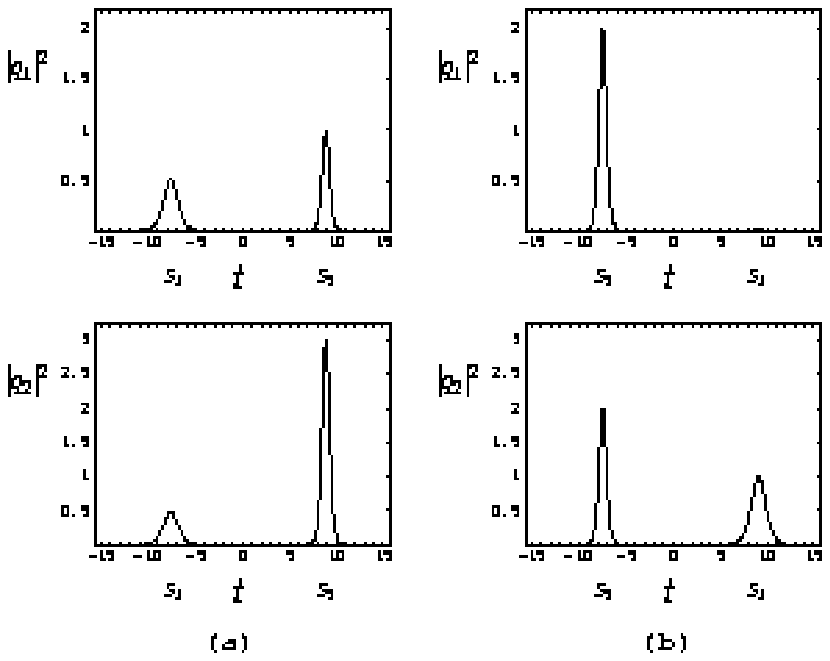, width=0.9\columnwidth}
\caption{Intensity profiles showing the collision scenario of two bright solitons of the mixed CNLS equations with $N=3$ in the absence of dark component, ($c_1=0$) with $\frac{\alpha_1^{(1)}}{\alpha_2^{(1)}} \neq \frac{\alpha_1^{(2)}}{\alpha_2^{(2)}}$.  (a) z=-4 and (b) z=4, given by the specific choice of parameters as given in the text.}
\label{3-a}
\end{figure}

\begin{figure}
\centering
\epsfig{figure=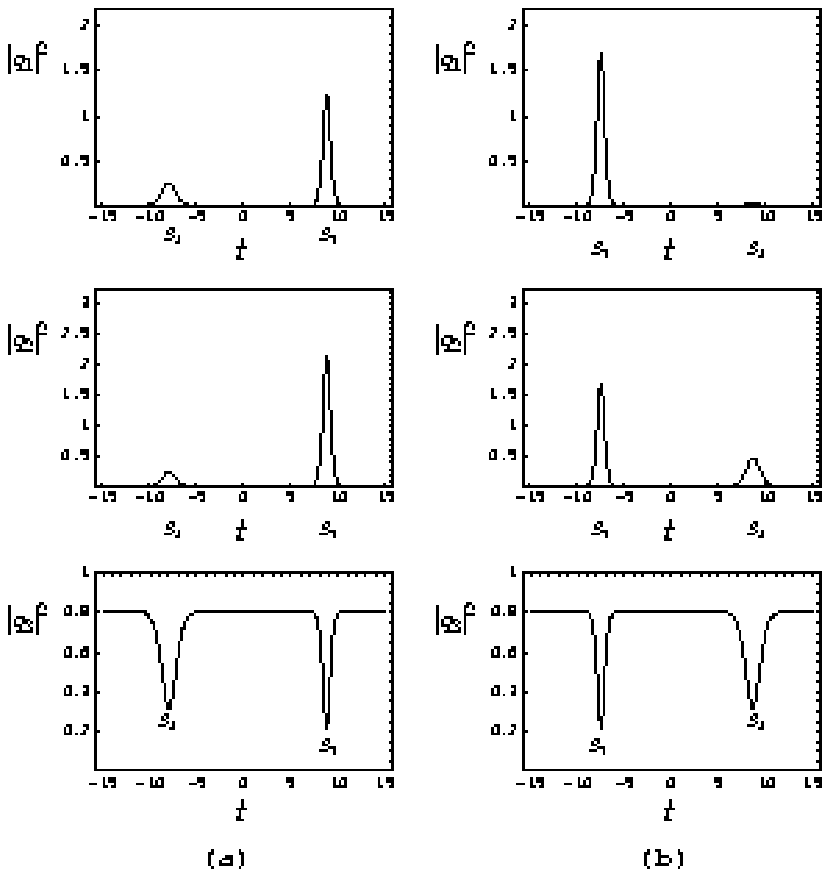, width=0.9\columnwidth}
\caption{Intensity profiles showing the collision scenario of two bright solitons of the mixed CNLS equations  with $N=3$ in the presence of dark component ($c_1\neq 0$) with$\frac{\alpha_1^{(1)}}{\alpha_2^{(1)}} \neq \frac{\alpha_1^{(2)}}{\alpha_2^{(2)}}$.  (a) z=-4 and (b) z=4, given by the specific choice of parameters as given in the text.}
\label{3-a-t}
\end{figure}

\begin{figure}
\centering
\epsfig{figure=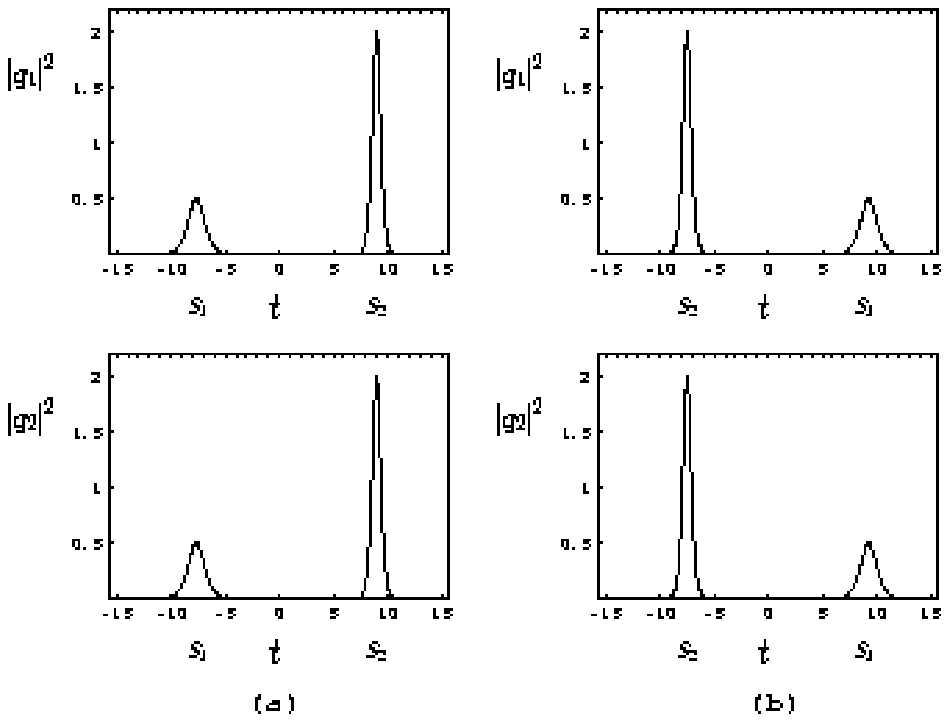, width=0.9\columnwidth}
\caption{Intensity profiles showing the collision scenario of two bright solitons of the mixed CNLS equations with $N=3$ in the absence of dark component ($c_1=0$), for $\frac{\alpha_1^{(1)}}{\alpha_2^{(1)}} = \frac{\alpha_1^{(2)}}{\alpha_2^{(2)}}$.  (a) z=-4 and (b) z=4.   The parameters are chosen as given in the text.}
\label{3}
\end{figure}

\begin{figure}
\centering
\epsfig{figure=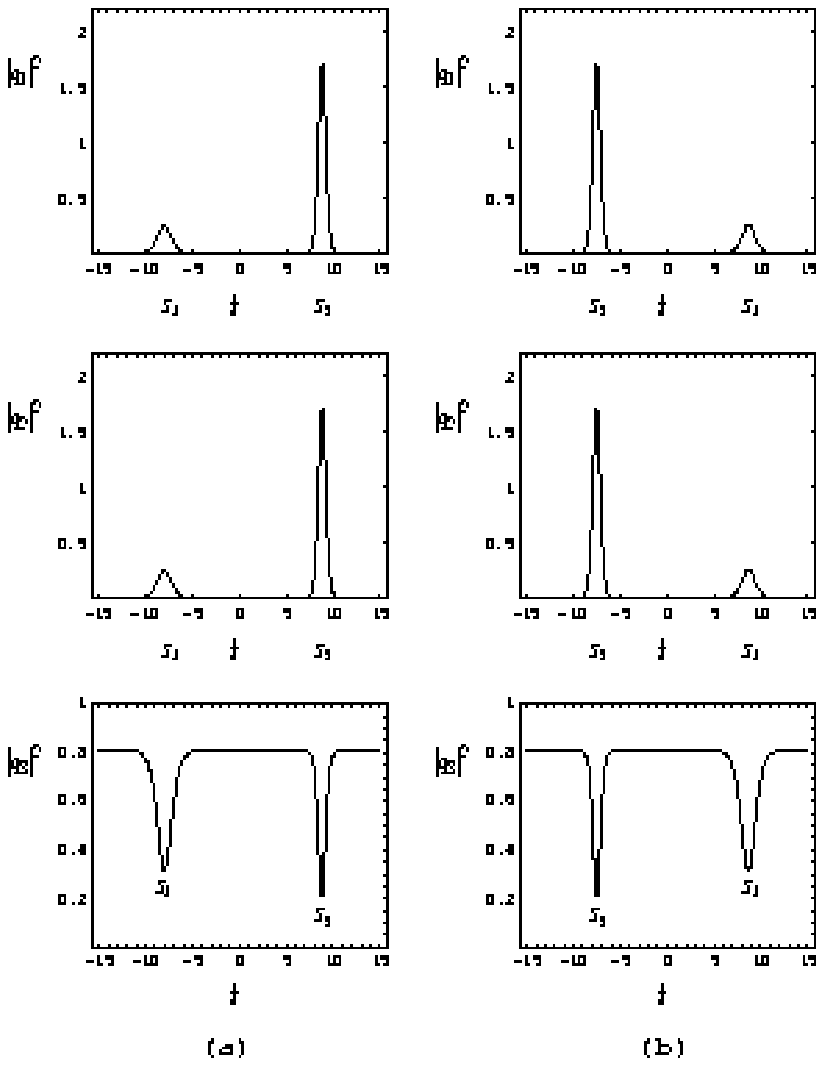, width=0.9\columnwidth}
\caption{Intensity profiles showing the collision scenario of two bright solitons of the mixed CNLS equations with $N=3$ in the presence of dark component ($c_1\neq 0$), for $\frac{\alpha_1^{(1)}}{\alpha_2^{(1)}} = \frac{\alpha_1^{(2)}}{\alpha_2^{(2)}}$.  (a) z=-4 and (b) z=4.  The parameters are chosen as in the text.}
\label{3-t}
\end{figure}

\begin{figure}
\centering
\epsfig{figure=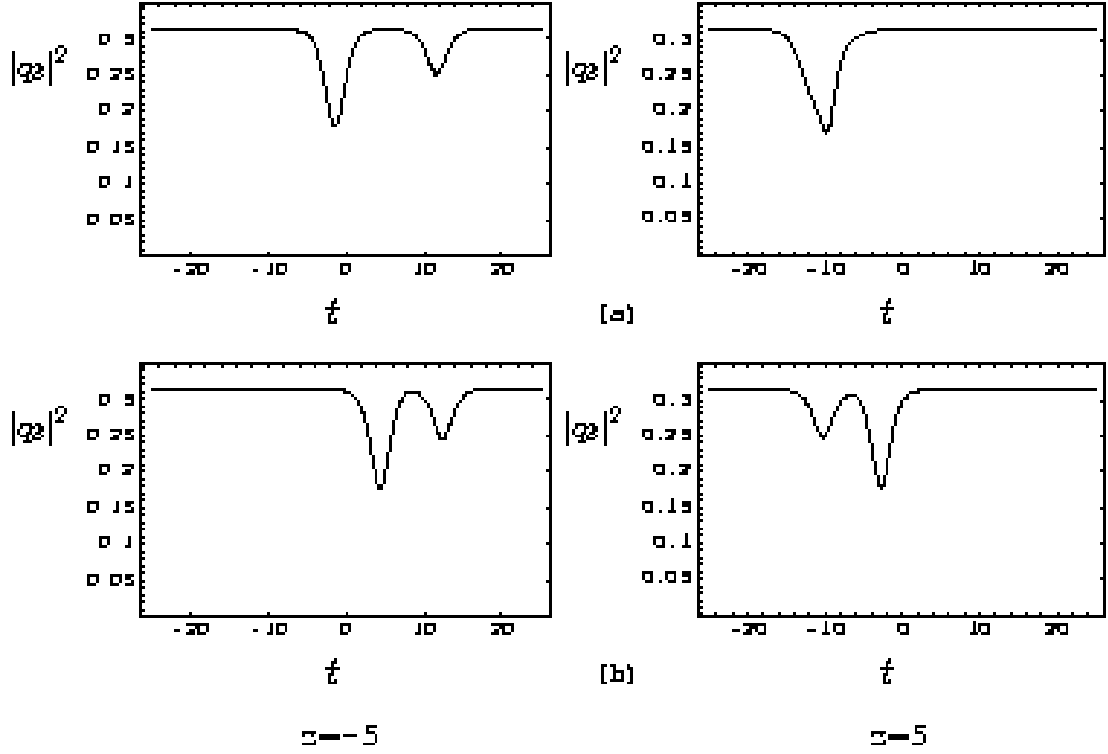, width=0.9\columnwidth}
\caption{Intensity profiles showing the propagation of two dark solitons in mixed CNLS system, before (z=-5) and after (z=5) collision: (a) for the special choice of parameters (as given in the text) with $\frac{\alpha_1^{(1)}}{\alpha_2^{(1)}} \neq \frac{\alpha_1^{(2)}}{\alpha_2^{(2)}}$; (b) for the special choice of parameters (as given in the text) with $\frac{\alpha_1^{(1)}}{\alpha_2^{(1)}} = \frac{\alpha_1^{(2)}}{\alpha_2^{(2)}}$.}
\label{sc}
\end{figure}

 We analyze two important physical quantities, namely (i) intensity and (ii) phase-shift of both bright and dark solitons.\\
\underline{(i) Intensities of bright and dark solitons:}\\
Analysing Eqs. (B1c), (B2c), (B3c) and (B4c),  we find that the amplitudes (intensities) of bright solitons before and after interaction are different.  In fact, the intensities of the colliding bright solitons before and after interactions can be related through the expression 
\bea
|A_j^{l+}|^2=|T_j^{l}|^2 |A_j^{l-}|^2, \quad j,l=1,2,
\eea
where the superscripts $l\pm$ represent the solitons designated as $S_1$ and $S_2$, at $z\rightarrow\pm\infty$. The transition intensities are identified from the Appendix B as 
\begin{subequations}
\bea
|T_j^{1}|^2&=&\frac{|1-\kappa_2(\alpha_2^{(j)}/\alpha_1^{(j)})|^2}{|1-\kappa_1\kappa_2|},\\
|T_j^{2}|^2&=&\frac{|1-\kappa_1\kappa_2|}{|1-\kappa_1(\alpha_2^{(j)}/\alpha_1^{(j)})|^2}, \quad j=1,2,\\
\kappa_1&=&\frac{\chi_{11}}{\chi_{21}}, \quad \kappa_2=\frac{\chi_{22}}{\chi_{12}},\\
\chi_{il}&=&\frac{(k_i+k_l^*)}{\sum_{j=1}^{2}(\alpha_i^{(j)}\alpha_l^{(j)*})}\left[1-\frac{|c_1|^2}{(k_i-ib_1)(k_l^*+ib_1)}\right],\quad i,l=1,2. 
\eea
\label{trans}
\end{subequations}
Note that the transition amplitudes $T_j^l$'s now are also functions of dark soliton parameters $c_1$ and $b_1$.  On the other hand, from Eqs. (\ref{2bq3-1}), (\ref{2bq3-2}), (\ref{2aq3-1}) and (\ref{2aq3-2}) we find that the intensities of dark solitons remain unchanged due to the  collision process.\\
(ii)\underline{Phase-shift of bright and dark solitons:}\\
The amplitude dependent phase-shift $\Phi_1\big(=\frac{R_3-R_2-R_1}{2}\big)$  for soliton $S_1$ can be expressed in the present case as
\begin{subequations}
\bea
\Phi_1=\frac{1}{2}\log\left[\frac{|k_1-k_2|^2}{(k_1+k_2^*)(k_2+k_1^*)}\left(1-U\left(\frac{(k_1+k_1^*)(k_2+k_2^*)}{(k_1+k_2^*)(k_2+k_1^*)}\frac{{\big[1-\frac{|c_1|^2}{P_1P_1^*}\big]}{\big[1-\frac{|c_1|^2}{P_2P_2^*}\big]}}{{\big[1-\frac{|c_1|^2}{P_1P_2^*}\big]}{\big[1-\frac{|c_1|^2}{P_2P_1^*}\big]}}\right)\right)\right],\nonumber\\
\label{p}
\eea
where
\bea
U=\frac{(\alpha_1^{(1)}\alpha_2^{(1)*}+\alpha_1^{(2)}\alpha_2^{(2)*})(\alpha_2^{(1)}\alpha_1^{(1)*}+\alpha_2^{(2)}\alpha_1^{(2)*})}{(|\alpha_1^{(1)}|^2+|\alpha_1^{(2)}|^2)(|\alpha_2^{(1)}|^2+|\alpha_2^{(2)}|^2)}.
\eea
\end{subequations}
Note that the soliton $S_2$ experiences an exactly opposite phase-shift $\Phi_2\; (=-\Phi_1)$.  The above phase-shift in turn results in the following change in relative separation distance between the solitons before and after collision, 
\bea
\Delta t_{12}&=&t_{12}^{-}-t_{12}^{+} =\frac{(k_{1R}+k_{2R})}{k_{1R}k_{2R}}\Phi_1.
\label{t}
\eea
The expression for $\Phi_1$ is given in Eq. (\ref{p}) clearly indicates that now the phase-shift depends on $\alpha$-parameters as well as $c_1$.  Thus the phase-shift and ultimately the relative separation distance between the solitons can be altered during a two-soliton collision process for a given combination of $k_j, c_1$ and $b_1$, by just varying the $\alpha$ parameters and as a whole the combined soliton profile gets altered. \\
This kind of collision scenario is shown in Fig. \ref{2b-1d}.  The corresponding intensity plots  show that the bright solitons undergo shape changing collisions characterized by intensity redistribution, amplitude dependent phase-shift and relative separation distance for the parametric choice $k_1=1+i, k_2=2-i, b_1=0.2, |c_1|=0.56, \alpha_1^{(1)}=1, \alpha_1^{(2)}=(32+i80)/89, \alpha_2^{(1)}=1, \alpha_2^{(2)}=1$.  The solitons (say $S_1$ and $S_2$) are well separated before and after collision in both the components $q_1$ and $q_2$.  In the $q_1$ component  the intensity of soliton $S_1$ gets suppressed while that of soliton $S_2$ is enhanced after interaction, whereas in the $q_2$ component it gets reversed.  Such shape changing collisions occur for $\frac{\alpha_1^{(1)}}{\alpha_2^{(1)}}\neq\frac{\alpha_1^{(2)}}{\alpha_2^{(2)}}$, which is quite general.  But when we choose $\frac{\alpha_1^{(1)}}{\alpha_2^{(1)}}=\frac{\alpha_1^{(2)}}{\alpha_2^{(2)}}$, the two solitons exhibit elastic collision.  It is instructive to note that although the dark solitons appear in the $q_3$ component, they indirectly influence the shape changing collisions through the carrier wave background parameters $c_1$ and $b_1$.  From the asymptotic analysis it follows that these background parameters influence the intensities of the colliding bright solitons before and after collision by different amounts through their explicit appearance in the transition intensities (see Eq. (\ref{trans})).  However the nature of the collision is unaltered.  We present below a detailed discussion to get a clear picture about the influence of dark solitons on bright soliton collision and vice-versa.\\
\underline{(a) Effect of dark soliton on the intensity of bright soliton:}\\
For a better understanding, we present the shape changing collision of bright solitons (i) in the absence of dark component, that is $c_1=0$, in Fig. \ref{3-a} and (ii) in the presence of dark component, that is $c_1\neq 0$, in Fig. \ref{3-a-t} for the above mentioned parametric choices.  From these figures we observe that, in the presence of dark component, the intensities of solitons $S_1$ in the $q_1$ component is decreased (increased) before(after) collision (as compared with Fig. \ref{3-a}), but not by the same amount.  Thus the effect of the dark component on the intensity of the bright soliton before its collision is different from the effect on the intensity after the collision.  But in the $q_2$ component there occurs decrement of intensity in both the solitons before and after collision due to the presence of dark soliton.  This confirms the fact that the presence of dark solitons indeed influences the cross phase coupling between the two components $q_1$ and $q_2$ which in turn affects the energy redistribution between those components as observed from figures.\\
However, the nature of collision, that is enhancement (suppression) in $S_1$ and suppression (enhancement) in $S_2$ in $q_1(q_2)$ component during shape changing collision process, is still preserved.  We also notice that in the case of standard elastic collision process resulting for the choice $\frac{\alpha_1^{(1)}}{\alpha_2^{(1)}}=\frac{\alpha_1^{(2)}}{\alpha_2^{(2)}}$, the role of the $c_1$ parameter on the amplitudes of colliding bright solitons before and after collision is same.  This is shown in Fig. \ref{3} and Fig. \ref{3-t}, for $c_1=0$ and $c_1 \neq 0$ respectively.  From the figures we observe that the intensities of bright solitons before and after collision are affected by the dark component by the same amount.\\
\underline{(b) Collision behaviour of dark solitons in the presence of bright solitons:}\\
The intensities of dark solitons in the $q_3$ component are unaffected  during the collision in the presence of the bright solitons in the $q_1$ and $q_2$ components.  This is obvious from the analytic expressions (\ref{2bq3-1}), (\ref{2bq3-2}), (\ref{2aq3-1}) and (\ref{2aq3-2}).  This type of collision scenario is shown in the third figure of Fig. \ref{2b-1d} (see also Fig. \ref{3-a-t}).   The analysis presented in Appendix B reveals the fact that the $\alpha$ parameters do affect the interaction of dark solitons through the shift in central position of the solitons which ultimately changes the separation distance between them after collision.  This phase-shift and the resulting change in the relative separation distance between the solitons can be obtained from Eq. (\ref{p}) and Eq. (\ref{t}), respectively.  In fact, the change in relative separation distance becomes more significant and displays interesting propagation and collision dynamics of solitons when the soliton velocities are moderately different and $k_{jR}$'s are equal.  For illustrative purpose, we consider the propagation of such composite two dark solitons arising for the choice $k_1=0.6-i, k_2=0.6-0.5i, b_1=0.2, |c_1|=0.56, \alpha_1^{(1)}=1, \alpha_1^{(2)}=i, \alpha_2^{(1)}=(22/55)-45i, \alpha_2^{(2)}=1$ in Fig. \ref{sc}, at $z=-5$ and $z=5$.  The analytic expression corresponding to the above choice $\left(k_{1R}=k_{2R}, \frac{\alpha_1^{(1)}}{\alpha_2^{(1)}}\neq\frac{\alpha_1^{(2)}}{\alpha_2^{(2)}}\right)$ is given by
\begin{subequations}
\bea
q_3=c_1 e^{i\zeta_1} \left[\frac{e^{\frac{Q_3}{2}} \mbox{cosh}(A+\frac{Q_3}{2})+e^{\frac{Q_{11}+Q_{22}}{2}} \mbox{cosh}(B+\frac{Q_{22}-Q_{11}}{2})+e^{\frac{Q_{12}+Q_{21}}{2}} \mbox{cos}(C+\frac{Q_{21}-Q_{12}}{2})}{e^{\frac{R_3}{2}}\mbox{cosh}(A+\frac{R_3}{2})+e^{\frac{R_{1}+R_{2}}{2}} \mbox{cosh}(B+\frac{R_{2}-R_{1}}{2})+e^{\frac{\delta_0^*+\delta_0}{2}} \mbox{cos}(C+\frac{\delta_0^*-\delta_0}{2})} \right],
\nonumber\\
\label{analsc}
\eea
where
\bea
A&=&2 k_{1R}[t-z(k_{1I}+k_{2I})],\\ 
B&=&2 k_{1R}[z(k_{1I}-k_{2I})],\\ 
C&=&(k_{2I}-k_{1I})t+(k_{1I}^2-k_{2I}^2)z.
\eea
\end{subequations}
Such a soliton solution can be viewed as soliton complex \cite{ref25,ref26} and all the parameters in Eq. (\ref{analsc}) are defined in Eq. (15).  Fig. \ref{sc}(a) shows that the dark soliton complex varies its profile during propagation within a finite distance.  This is a consequence of change in the relative separation distance  due to the presence of $\alpha$-parameters.   However for the same choice of $k_j$'s, $b_1$, and $c_1$ but with $\frac{\alpha_1^{(1)}}{\alpha_2^{(1)}}=\frac{\alpha_1^{(2)}}{\alpha_2^{(2)}}$,  the dark soliton complex does not vary its profile after propagation through the same distance.  This is shown in Fig. \ref{sc}(b).  This kind of behaviour is a striking feature of multisoliton complexes \cite{ref25,ref26}.  The bright counterparts of this dark soliton also form bright soliton complexes of variable shape and such bright soliton complexes in CNLS equations with focusing nonlinearity are discussed in Refs. \cite{ref13,ref25,ref26,ref27}.  Work is in progress on such bright-dark multisoliton complexes. \\ 
\subsubsection{1-Bright$-$2-Dark Soliton Collision:} 
Two soliton solution given by Eq. (17) describes the collision of solitons in which bright soliton collision takes place in the $q_1$ component and the dark soliton collision occurs in the $q_2$ and $q_3$ components.  The asymptotic expressions for the colliding solitons before and after collision are given in Appendix B. Equations (B5), (B6), (B7) and (B8)  show that the intensities of bright solitons before and after collisions are same , (i.e.) $|A_1^{l-}|=|A_1^{l+}|$, $l=1,2$.  Similarly the intensities of the dark solitons are also unaltered after collision.  This indicates that the bright as well as the dark solitons undergo elastic collision.  However there occurs a dark soliton parameter dependent phase-shift due to collision.  The phase-shift of soliton $S_1$ in both the bright and dark components is given by
\bea
\Phi_1&=&\bigg( \frac{R_3-R_2-R_1}{2}\bigg),\nonumber \\
&=&\frac{1}{2}\log\left(\frac{|k_1-k_2|^2}{(k_1+k_2^*)^2(k_2+k_1^*)^2} \frac{\left|1+\sum_{v=1}^2\frac{|c_v|^2}{(k_1-ib_v)(k_2-ib_v)}\right|^2}{\left|1-\sum_{v=1}^2\frac{|c_v|^2}{(k_1-ib_v)(k_2-ib_v)}\right|^2}\right).
\eea
Similarly soliton $S_2$ undergoes a phase-shift $\Phi_2=-\Phi_1$.  The dark component parameters $c_1, c_2, b_1$ and $b_2$ influence the phase-shift whereas the bright component parameter $\alpha$ does not alter the phase-shift.

The above asymptotic analysis of two soliton solution of 2-CNLS and 3-CNLS can be extended straight forwardly to N-CNLS equations, with arbitrary N.  By generalizing the above study we also point out that for mixed N-CNLS equations (1), the bright and dark solitons can be split up in $(N-1)$ ways, starting from (N-1)-bright$-$1-dark soliton solution, (N-2)-bright$-$2-dark soliton solution, up to 1-bright$-$(N-1)-dark soliton solution.  We also arrive at an important conclusion that for $m$-bright$-$$n$-dark two-soliton solution, where $m+n=N$, the shape changing collision of bright solitons takes place only when $m\geq2$.   
\section{CONCLUSION}
To conclude, we have obtained the explicit mixed type (bright-dark) soliton solutions for the multicomponent mixed coupled nonlinear Schr{\"o}dinger equations using Hirota's bilinearization method.  In particular, we have shown that these coupled bright and dark solitons possess rich structure and become more general than individual bright/dark solitons.  Next, our analysis of their collision behaviour reveals the fact that there exist elastic collisions of bright as well as dark solitons for the two component case.  We observe that for this case the bright solitons are affected uniformly by the dark solitons.  The bright soliton parameters $\alpha$'s have no effect on the phase-shift whereas it is influenced by the dark soliton parameters $c_1$ and $b_1$.  The important observation of this study is that for more than two components, if the bright solitons appear in at least two components, then those bright solitons undergo shape changing collisions characterized by intensity redistribution, amplitude dependent phase-shift and relative separation distances, but their counterpart (dark solitons) exhibits only elastic collisions but with amplitude dependent phase-shift.  This identification can find potential applications in optical as well as matter wave switching devices where the switching is performed through shape changing collision of solitons.  Further, in contrast to the $N$ component Manakov system here the phase-shift of bright as well as dark solitons is characterized by dark soliton parameters  $c_1$ and $b_1$ in addition to the $\alpha$ parameters. Also, we observe that the dark soliton parameter $c_1$ influences the intensity of bright solitons by different amount when the bright soliton parameters $\alpha$'s are such that $\frac{\alpha_1^{(1)}}{\alpha_2^{(1)}}\neq\frac{\alpha_1^{(2)}}{\alpha_2^{(2)}}$ while their amplitudes are affected by dark solitons by the same amount, for $\frac{\alpha_1^{(1)}}{\alpha_2^{(1)}}=\frac{\alpha_1^{(2)}}{\alpha_2^{(2)}}$.  One more noticeable observation is that the dark solitons  vary their profiles depending on $\alpha$ parameters during propagation as in multisoliton complexes \cite{ref26,ref27}.  The various results obtained from the study will give further insight into the bright-dark paired solitons, soliton complexes formation, collision in Boson-Fermion mixtures and in nonlinear left-handed materials and their applications in switching devices. 
\section*{ACKNOWLEDGEMENTS}
T. K. acknowledges the support of Department of Science and Technology, Government of India under the DST Fast Track Project for Young Scientists.  He is also grateful to the constant support of the Principal and Management of Bishop Heber College, Tiruchirapalli.  The work of M. V. and M. L. are supported by a DST-IRPHA project.  M. L. is also supported by a DST Ramanna Fellowship. 

\appendix
\section{Asymptotic analysis of bright-dark two-soliton solution of mixed 2-CNLS equations}
In the limit $z\rightarrow\pm\infty$ the two-soliton solution (11) takes the following asymptotic forms for the choice $k_{1R}, k_{2R}>0\; \mbox{and}\; k_{1I}>k_{2I}$.  For other choices of $k_i$'s, $i=1,2,$ similar analysis can be made.\\
\underline { {\bf {a)}} Before collision (limit $z\rightarrow-\infty$):}\\ 
 (i)  Soliton 1 ($\eta_{1R}\simeq0, \eta_{2R}\rightarrow -\infty$):
\begin{subequations}
\bea
q_1&\simeq& A_1^{1-} k_{1R} e^{i \eta_{1I}} \mbox{sech}\bigg(\eta_{1R}+\frac{R_1}{2}\bigg),\\
q_2&\simeq& -c_1\;e^{i(\zeta_1+\phi_1^{(1)})} \bigg[\mbox{cos}\phi_1^{(1)} \mbox{tanh}\bigg(\eta_{1R}+\frac{R_1}{2}\bigg)+i \mbox{sin}\phi_1^{(1)}\bigg],\\
\hspace{-2cm}\mbox{where}\qquad \qquad \qquad \qquad  \nonumber\\
A_1^{1-}&=&\bigg(\frac{\alpha_1^{(1)}}{\alpha_1^{(1)*}}\bigg)^\frac{1}{2} \frac{1}{k_{1R}} \sqrt{k_{1R}^2-|c_1|^2 \mbox{cos}^2\phi_1^{(1)}}, \\
\phi_1^{(1)}&=& \mbox{tan}^{-1}\bigg(\frac{k_{1I}-b_1}{k_{1R}}\bigg).
\label{phi1}
\eea
\label{2-s1}
\end{subequations}
\\
(ii)  Soliton 2 ($\eta_{2R}\simeq0, \eta_{1R}\rightarrow \infty$):
\begin{subequations}
\bea
q_1&\simeq& A_1^{2-} k_{2R} e^{i \eta_{2I}} \mbox{sech}\bigg(\eta_{2R}+\frac{(R_3-R_1)}{2}\bigg),\\
q_2&\simeq& c_1\;e^{i(\zeta_1+\phi_2^{(1)})}\frac{(k_1-ib_1)}{(k_1^*+ib_1)} \bigg[\mbox{cos}\phi_2^{(1)} \mbox{tanh}\bigg(\eta_{2R}+\frac{(R_3-R_1)}{2}\bigg)+i \mbox{sin}\phi_2^{(1)}\bigg],\\
\hspace{-2cm}\mbox{where}\qquad \quad \nonumber\\
A_1^{2-}&=&\bigg(\frac{\alpha_2^{(1)}}{\alpha_2^{(1)*}}\bigg)^\frac{1}{2}\frac{1}{k_{2R}} \sqrt{k_{2R}^2-|c_1|^2 \mbox{cos}^2\phi_2^{(1)}}\bigg(\frac{X}{X^*}\bigg)\\
\hspace{-2cm}\mbox{and}\qquad \qquad  \nonumber\\
\phi_2^{(1)}&=&\mbox{tan}^{-1}\bigg(\frac{k_{2I}-b_1}{k_{2R}}\bigg).
\label{phi2}
\eea
In the above expressions,
\bea
X=&&(k_1-k_2)(k_1+k_2^*)(k_1^*+ib_1)([(k_1-ib_1)(k_2^*+ib_1)-|c_1|^2][(k_1-ib_1)(k_2-ib_1)+|c_1|^2])^\frac{1}{2}.\nonumber
\eea
\end{subequations}
\underline {{\bf {b)}} After collision (limit $z\rightarrow\infty$):}\\ 
(i)  Soliton 1 ($\eta_{1R}\simeq0, \eta_{2R}\rightarrow \infty$):
\begin{subequations}
\bea
q_1&\simeq& A_1^{1+} k_{1R} e^{i \eta_{1I}} \mbox{sech}\bigg(\eta_{1R}+\frac{(R_3-R_2)}{2}\bigg),\\
q_2&\simeq& c_1\;e^{i(\zeta_1+\phi_1^{(1)})}\frac{(k_2-ib_1)}{(k_2^*+ib_1)}\bigg[\mbox{cos}\phi_1^{(1)} \mbox{tanh}\bigg(\eta_{1R}+\frac{(R_3-R_2)}{2}\bigg)+i \mbox{sin}\phi_1^{(1)}\bigg],\\
\hspace{-2cm}\mbox{where}\qquad \quad  \nonumber\\
A_1^{1+}&=&\bigg(\frac{\alpha_1^{(1)}}{\alpha_1^{(1)*}}\bigg)^\frac{1}{2}\frac{1}{k_{1R}} \sqrt{k_{1R}^2-|c_1|^2 \mbox{cos}^2\phi_1^{(1)}}\bigg(\frac {Y}{Y^*}\bigg).
\eea
Here
\bea
Y=&&(k_1-k_2)(k_2+k_1^*)(k_2^*+ib_1)([(k_1-ib_1)(k_2-ib_1)+|c_1|^2][(k_1^*+ib_1)(k_2-ib_1)-|c_1|^2])^\frac{1}{2} .\nonumber
\eea
\end{subequations}
(ii)  Soliton 2 ($\eta_{2R}\simeq0, \eta_{1R}\rightarrow -\infty$):
\begin{subequations}
\bea
q_1&\simeq& A_1^{2+} k_{2R} e^{i \eta_{2I}} \mbox{sech}\bigg(\eta_{2R}+\frac{R_2}{2}\bigg),\\
q_2&\simeq& - c_1\;e^{i(\zeta_1+\phi_2^{(1)})}\bigg[\mbox{cos}\phi_2^{(1)} \mbox{tanh}\bigg(\eta_{2R}+\frac{R_2}{2}\bigg)+i \mbox{sin}\phi_2^{(1)}\bigg],\\
\hspace{-4cm}\mbox{where}\qquad  \qquad \qquad\nonumber\\
A_1^{2+}&=&\bigg(\frac{\alpha_2^{(1)}}{\alpha_2^{(1)*}}\bigg)^\frac{1}{2} \frac{1}{k_{2R}} \sqrt{k_{2R}^2-|c_1|^2 \mbox{cos}^2\phi_2^{(1)}}.
\eea
\label{2-s2}
\end{subequations}
Note that in the above expressions though $A_j^{l+}\neq A_j^{l-}, \; j,l=1,2,\; |A_j^{l+}|= |A_j^{l-}|$ and thereby confirming that no intensity redistribution occurs in the case of bright soliton and also for dark soliton (see Eqs. (A1b), (A2b), (A3b) and (A4b)).  The phase-shift $\Phi_1$ ($\Phi_2$) of soliton $S_1$ ($S_2$) in the bright and dark component obtained from the above asymptotic expressions is given by Eq. (20a).
\section{Asymptotic analysis of bright-dark two-soliton solution of mixed 3-CNLS equations}
\subsection{2-bright$-$1-dark soliton solution}
Here also we assume $k_{1R}, k_{2R}>0\; \mbox{and}\; k_{1I}>k_{2I}$, then the two soliton  (2-bright$-$1-dark) solution (15) takes the following forms asymptotically ($z\rightarrow\pm\infty$).\\ 
\underline {{\bf {a)}} Before collision (limit $z\rightarrow-\infty$):}\\ 
(i)  Soliton 1 ($\eta_{1R}\simeq0, \eta_{2R}\rightarrow -\infty$):
\begin{subequations}
\bea
\left(
\begin{array}{c}
  q_1 \\\\
  q_2
\end{array}
\right)&&\simeq \left(
\begin{array}{c}
 A_{1}^{1-} \\\\
 A_{2}^{1-}  \\ 
\end{array}
\right) k_{1R} \mbox{\mbox{sech}}\left(\eta_{1R}+\frac{R_1}{2}
\right)e^{i\eta_{1I}},\\
q_3&&\simeq -c_1\;e^{i(\zeta_1+\phi_1^{(1)})} \bigg[\mbox{cos}\phi_1^{(1)} \mbox{tanh}\bigg(\eta_{1R}+\frac{R_1}{2}\bigg)+i \mbox{sin}\phi_1^{(1)}\bigg],
\label{2bq3-1}
\eea
where
\bea
\left(
\begin{array}{c}
A_{1}^{1-} \\\\
A_{2}^{1-}
\end{array}
\right)\simeq \left(
\begin{array}{c}
\alpha_1^{(1)}\\\\
\alpha_1^{(2)}
\end{array}
\right)  \frac{\sqrt{k_{1R}^2-|c_1|^2 \mbox{cos}^2\phi_1^{(1)}}}{k_{1R}\big(|\alpha_1^{(1)}|^2+|\alpha_1^{(2)}|^2\big)^\frac{1}{2}}.
\eea
\label{s1}
\end{subequations}
(ii)  Soliton 2 ($\eta_{2R}\simeq0, \eta_{1R}\rightarrow \infty$):
\begin{subequations}
\bea
\left(
\begin{array}{c}
  q_1 \\\\
  q_2
\end{array}
\right)&&\simeq \left(
\begin{array}{c}
 A_{1}^{2-} \\\\
 A_{2}^{2-} 
\end{array}
\right) k_{2R} \mbox{\mbox{sech}}\left(\eta_{2R}+\frac{(R_3-R_1)}{2}
\right)e^{i\eta_{2I}},\\
q_3&&\simeq c_1\;e^{i(\zeta_1+\phi_2^{(1)})}\frac{(k_1-ib_1)}{(k_1^*+ib_1)} \bigg[\mbox{cos}\phi_2^{(1)} \mbox{tanh}\bigg(\eta_{2R}+\frac{(R_3-R_1)}{2}\bigg)+i \mbox{sin}\phi_2^{(1)}\bigg],\nonumber\\
\label{2bq3-2}
\eea
where
\bea
\left(
\begin{array}{c}
A_{1}^{2-} \\\\
A_{2}^{2-}
\end{array}
\right)\simeq \left(
\begin{array}{c}
\alpha_2^{(1)}\bigg(\frac{1-(\alpha_1^{(1)}/\alpha_2^{(1)})\kappa_1}{\sqrt{1-\kappa_1\kappa_2}}\bigg)\\\\
\alpha_2^{(2)}\bigg(\frac{1-(\alpha_1^{(2)}/\alpha_2^{(2)})\kappa_1}{\sqrt{1-\kappa_1\kappa_2}}\bigg)
\end{array}
\right)\left(\frac{(k_2-k_1)|k_1+k_2^*|}{(k_2+k_1^*)|k_1-k_2|} \frac{\sqrt{k_{2R}^2-|c_1|^2 \mbox{cos}^2\phi_2^{(1)}}}{k_{2R}\big(|\alpha_2^{(1)}|^2+|\alpha_2^{(2)}|^2\big)^\frac{1}{2}}\right).
\eea
\end{subequations}
\underline {{\bf {b)}} After collision (limit $z\rightarrow\infty$):}\\ 
(i)  Soliton 1 ($\eta_{1R}\simeq0, \eta_{2R}\rightarrow \infty$):
\begin{subequations}
\bea
\left(
\begin{array}{c}
  q_1 \\\\
  q_2
\end{array}
\right)&&\simeq \left(
\begin{array}{c}
 A_{1}^{1+} \\\\
 A_{2}^{1+} 
\end{array}
\right) k_{1R} \mbox{\mbox{sech}}\left(\eta_{1R}+\frac{(R_3-R_2)}{2}
\right)e^{i\eta_{1I}},\\
q_3&&\simeq c_1\;e^{i(\zeta_1+\phi_1^{(1)})}\frac{(k_2-ib_1)}{(k_2^*+ib_1)}  \bigg[\mbox{cos}\phi_1^{(1)} \mbox{tanh}\bigg(\eta_{1R}+\frac{(R_3-R_2)}{2}\bigg)+i \mbox{sin}\phi_1^{(1)}\bigg],\nonumber\\
\label{2aq3-1}
\eea
where
\bea
\left(
\begin{array}{c}
A_{1}^{1+} \\\\
A_{2}^{1+}
\end{array}
\right)\simeq \left(
\begin{array}{c}
\alpha_1^{(1)}\bigg(\frac{1-(\alpha_2^{(1)}/\alpha_1^{(1)})\kappa_2}{\sqrt{1-\kappa_1\kappa_2}}\bigg)\\\\
\alpha_1^{(2)}\bigg(\frac{1-(\alpha_2^{(2)}/\alpha_1^{(2)})\kappa_2}{\sqrt{1-\kappa_1\kappa_2}}\bigg)
\end{array}
\right)\left(\frac{(k_1-k_2)|k_1+k_2^*|}{(k_1+k_2^*)|k_1-k_2|} \frac{\sqrt{k_{1R}^2-|c_1|^2 \mbox{cos}^2\phi_1^{(1)}}}{k_{1R}\big(|\alpha_1^{(1)}|^2+|\alpha_1^{(2)}|^2\big)^\frac{1}{2}}\right).
\eea
\end{subequations}
(ii)  Soliton 2 ($\eta_{2R}\simeq0, \eta_{1R}\rightarrow -\infty$):
\begin{subequations}
\bea
\left(
\begin{array}{c}
  q_1 \\\\
  q_2
\end{array}
\right)&&\simeq \left(
\begin{array}{c}
 A_{1}^{2+} \\\\
 A_{2}^{2+} 
\end{array}
\right) k_{2R} \mbox{\mbox{sech}}\left(\eta_{2R}+\frac{R_2}{2}
\right)e^{i\eta_{2I}},\\
q_3&&\simeq -c_1\;e^{i(\zeta_1+\phi_2^{(1)})}\bigg[\mbox{cos}\phi_2^{(1)} \mbox{tanh}\bigg(\eta_{2R}+\frac{R_2}{2}\bigg)+i \mbox{sin}\phi_2^{(1)}\bigg],
\label{2aq3-2}
\eea
where
\bea
\left(
\begin{array}{c}
A_{1}^{2+} \\\\
A_{2}^{2+}
\end{array}
\right)\simeq \left(
\begin{array}{c}
\alpha_2^{(1)}\\\\
\alpha_2^{(2)}
\end{array}
\right) \frac{\sqrt{k_{2R}^2-|c_1|^2 \mbox{cos}^2\phi_2^{(1)}}}{k_{2R}\big(|\alpha_2^{(1)}|^2+|\alpha_2^{(2)}|^2\big)^\frac{1}{2}} .
\eea
\label{s2}
\end{subequations}
In the above equations $\phi_1^{(1)}$ and $\phi_2^{(1)}$ are defined in Eqs. (\ref{phi1}) and (\ref{phi2}), respectively.  Note that $A_j^{l+}\neq A_j^{l-}$, and also in general the intensities $|A_j^{l+}|^2\neq |A_j^{l-}|^2,\; j,l=1,2$, except when $\frac{\alpha_1^{(1)}}{\alpha_2^{(1)}} = \frac{\alpha_1^{(2)}}{\alpha_2^{(2)}}$.  The explicit relation between the intensities before and after collision is given by Eq. (\ref{trans}) in the text.  The phase-shifts $\Phi_1$ and $\Phi_2$ suffered by solitons $S_1$ and $S_2$, respectively are given in Eq. (23a).  

\subsection{1-bright$-$2-dark soliton solution}
Considering the 1-bright$-$2-dark two soliton solution (17), the analysis in the asymptotic limits can be performed as follows (with $k_{1R}$, $k_{2R}>0$ and $k_{1I}$$>$$k_{2I}$).\\
\underline{ {\bf {a)}} Before collision (limit $z\rightarrow-\infty$):}\\ 
(i)  Soliton 1 ($\eta_{1R}\simeq0, \eta_{2R}\rightarrow -\infty$):
\begin{subequations}
\bea
q_1\simeq&& A_1^{1-} k_{1R} e^{i \eta_{1I}} \mbox{sech}\bigg(\eta_{1R}+\frac{R_1}{2}\bigg),\\
q_{j+1}\simeq && -c_j\;e^{i(\zeta_j+\phi_1^{(j)})}\bigg[\mbox{cos}\phi_1^{(j)} \mbox{tanh}\bigg(\eta_{1R}+\frac{R_1}{2}\bigg)+i \mbox{sin}\phi_1^{(j)}\bigg], \quad j=1,2,
\eea
where
\bea
A_1^{1-}=&&\bigg(\frac{\alpha_1^{(1)}}{\alpha_1^{(1)*}}\bigg)^\frac{1}{2} \frac{1}{k_{1R}} \sqrt{k_{1R}^2-\left(|c_1|^2 \mbox{cos}^2\phi_1^{(1)}+|c_2|^2 \mbox{cos}^2\phi_1^{(2)}\right)},\\
\phi_1^{(j)}=&&\mbox{tan}^{-1}\bigg(\frac{k_{1I}-b_j}{k_{1R}}\bigg).
\eea
\end{subequations}
(ii)  Soliton 2 ($\eta_{2R}\simeq0, \eta_{1R}\rightarrow \infty$):
\begin{subequations}
\bea
q_1\simeq && A_1^{2-} k_{2R} e^{i \eta_{2I}} \mbox{sech}\bigg(\eta_{2R}+\frac{(R_3-R_1)}{2}\bigg), \\
q_{j+1}\simeq&& c_j\;e^{i(\zeta_j+\phi_2^{(j)})}\frac{(k_1-ib_j)}{(k_1^*+ib_j)} \bigg[\mbox{cos}\phi_2^{(j)} \mbox{tanh}\bigg(\eta_{2R}+\frac{(R_3-R_1)}{2}\bigg)+i \mbox{sin}\phi_2^{(j)}\bigg],\nonumber\\
&&\qquad \qquad \qquad \qquad \qquad \qquad \qquad \qquad \qquad \qquad \qquad \qquad j=1,2,
\eea
where
\bea
A_1^{2-}=&&A_1^{2+}\left(\frac{Q_1Q_2}{Q_1^*Q_2^*}\right),\quad \phi_2^{(j)}=\mbox{tan}^{-1}\bigg(\frac{k_{2I}-b_j}{k_{2R}}\bigg),\\ 
Q_1=&&(k_1-k_2)\left[1+\frac{|c_1^2|}{(k_1-ib_1)(k_2-ib_1)}+\frac{|c_2^2|}{(k_1-ib_2)(k_2-ib_2)}\right]^{1/2},\\
Q_2=&&(k_1+k_2^*)\left[1-\frac{|c_1^2|}{(k_1-ib_1)(k_2^*+ib_1)}+\frac{|c_2^2|}{(k_1-ib_2)(k_2^*+ib_2)}\right]^{1/2}.   
\eea
\end{subequations}
Here $A_1^{2+}$ is defined in Eq. (\ref{a12+}) given below.\\
\underline { {\bf {b)}} After collision (limit $z\rightarrow\infty$):}\\ 
(i)  Soliton 1 ($\eta_{1R}\simeq0, \eta_{2R}\rightarrow \infty$):
\begin{subequations}
\bea
q_1\simeq && A_1^{1+} k_{1R} e^{i \eta_{1I}} \mbox{sech}\bigg(\eta_{1R}+\frac{(R_3-R_2)}{2}\bigg),\\
q_{j+1}\simeq&& c_j\;e^{i(\zeta_j+\phi_1^{(j)})}\frac{(k_2-ib_j)}{(k_2^*+ib_j)}\bigg[\mbox{cos}\phi_1^{(j)} \mbox{tanh}\bigg(\eta_{1R}+\frac{(R_3-R_2)}{2}\bigg)+i \mbox{sin}\phi_1^{(j)}\bigg],\nonumber\\
&&\qquad \qquad \qquad \qquad \qquad \qquad \qquad \qquad \qquad \qquad \qquad \qquad j=1,2,\\
&&\hspace {-3cm} \mbox{where}\nonumber\\
A_1^{1+}=&&A_1^{1-}\left(\frac{Q_1Q_2^*}{Q_1^*Q_2}\right).
\eea
\end{subequations}
(ii)  Soliton 2 ($\eta_{2R}\simeq0, \eta_{1R}\rightarrow -\infty$):
\begin{subequations}
\bea
q_1\simeq&& A_1^{2+} k_{2R} e^{i \eta_{2I}} \mbox{sech}\bigg(\eta_{2R}+\frac{R_2}{2}\bigg), \\
q_{j+1}\simeq&& -c_j\;e^{i(\zeta_j+\phi_2^{(j)})}\bigg[\mbox{cos}\phi_2^{(j)} \mbox{tanh}\bigg(\eta_{2R}+\frac{R_2}{2}\bigg)+i \mbox{sin}\phi_2^{(j)}\bigg],\quad j=1,2,\\
&&\hspace {-3.5cm} \mbox{where}\nonumber\\
A_1^{2+}=&&\bigg(\frac{\alpha_2^{(1)}}{\alpha_2^{(1)*}}\bigg)^\frac{1}{2} \frac{1}{k_{2R}} \sqrt{k_{2R}^2-\left(|c_1|^2 \mbox{cos}^2\phi_2^{(1)}+|c_2|^2 \mbox{cos}^2\phi_2^{(2)}\right)}.
\label{a12+}
\eea
\end{subequations}
Note that in the above expressions $|A_1^{l+}|= |A_1^{l-}|, l=1,2,$ and thereby confirming that no intensity redistribution occurs in the case of bright soliton, and also for dark solitons [see Eqs. (B5b), (B6b), (B7b) and (B8b)].




\end {document}